\documentclass{article}

\usepackage[preprint]{neurips_2026}

\usepackage[utf8]{inputenc}
\usepackage[T1]{fontenc}
\usepackage{hyperref}
\usepackage{url}
\usepackage{booktabs}
\usepackage{amsfonts}
\usepackage{amsmath}
\usepackage{nicefrac}
\usepackage{microtype}
\usepackage{xcolor}
\usepackage{graphicx}
\usepackage{multirow}
\usepackage{enumitem}
\usepackage{subcaption}
\usepackage{hyphenat}
\usepackage{tikz}
\usetikzlibrary{arrows.meta}
\usepackage[breakable]{tcolorbox}

\newtcolorbox{promptbox}[1][]{
  colback=gray!5,
  colframe=gray!50,
  fontupper=\small\ttfamily,
  breakable,
  left=6pt, right=6pt, top=4pt, bottom=4pt,
  title={\sffamily\bfseries #1},
  coltitle=black,
  colbacktitle=gray!15,
}

\title{Talk is (Not) Cheap: A Taxonomy and Benchmark Coverage Audit for LLM Attacks}

\author{%
  Karthik Raghu Iyer \\
  Palo Alto Networks \\
  \texttt{kiyer@paloaltonetworks.com} \\
  \And
  Yazdan Jamshidi \\
  Palo Alto Networks \\
  \texttt{yjamshidi@paloaltonetworks.com} \\
  \And
  Nicholas Bray \\
  Palo Alto Networks \\
  \texttt{nbray@paloaltonetworks.com} \\
  \And
  Alexey Shvets \\
  Palo Alto Networks \\
  \texttt{ashvets@paloaltonetworks.com} \\
}
\begin{document}

\maketitle

\begin{abstract}
We introduce a reusable framework for auditing whether LLM attack benchmarks collectively cover the threat surface: a 4$\times$6 Target $\times$ Technique matrix grounded in STRIDE, constructed from a 507-leaf taxonomy---401 data-populated and 106 threat-model-derived leaves---of inference-time attacks extracted from 932 arXiv security studies (2023--2026). The matrix enables benchmark-external validation---auditing collective coverage rather than individual benchmark consistency. Applying it to six public benchmarks reveals that the three primary frameworks (HarmBench, InjecAgent, AgentDojo) occupy non-overlapping cells covering at most 25\% of the matrix, while entire STRIDE threat categories (Service Disruption, Model Internals) lack any standardized evaluation, despite published attacks in these categories achieving 46$\times$ token amplification and 96\% attack success rates through mechanisms which no benchmark tests. The corpus of 2,521 unique attack groups further reveals pervasive naming fragmentation (up to 29 surface forms for a single attack) and heavy concentration in Safety \& Alignment Bypass, structural properties invisible at smaller scale. The taxonomy, attack records, and coverage mappings are released as extensible artifacts; as new benchmarks emerge, they can be mapped onto the same matrix, enabling the community to track whether evaluation gaps are closing.
\end{abstract}

\section{Introduction}
\label{sec:intro}

Large language models (LLMs) are deployed widely, but their safety alignment is fundamentally limited by training, and specific prompts can trigger latent harmful behavior~\citep{wolf2023fundamental}.

This landscape has grown since 2023, when it was discovered that optimization-based methods could compromise the safety alignments of LLMs, and produce both harmful and toxic content~\citep{zou2023universal}. These attacks can be embedded in websites, documents, or tool outputs, with consequences ranging from fraud and malware distribution to information exfiltration~\citep{greshake2023not}. As other forms of attack vehicles, such as prompt engineering and fine tuning, have been discovered, the attack landscape has only further increased and splintered~\citep{wei2024jailbroken, qi2023finetuning}. Our corpus of papers, spanning from H1 2023 to H1 2026, includes 932 unique papers with over 6{,}300 referenced attack mentions.

This rapid growth creates three interconnected problems:
\begin{enumerate}[leftmargin=*,itemsep=2pt]
    \item \emph{The field lacks a clear taxonomy focused on quantitative results.} Existing surveys and systematizations cover only 50--150 papers through manual review, which provide a valuable framework to build off of, but do not guarantee complete coverage~\citep{xu2025survey,kim2026attacklandscape,grosse2024practical}. An automated process with human review can ensure an analysis at a much higher scale, necessary as the number of papers in the field of LLM security research grows past five digits~\citep{kostikova2025lllms}.
    \item \emph{Naming fragmentation is pervasive and consistent throughout the attacks.} Without a clear standard by which attacks can be categorized, we see that attacks are often rediscovered with slight variations, leading to multiple surface forms and making reproducibility/meta-analyses more difficult. For example, GCG (Greedy Coordinate Gradient)~\citep{zou2023universal} appears under 29 distinct names across 376 papers in our corpus.
    \item \emph{Blind spots within evaluation infrastructure.} Without clearly understanding the attack landscape, blind spots within defense frameworks can be present, leaving users vulnerable to an undefined degree. To complete this, we adapt the meta-review methodology defined by \citet{liao2021we}, applying their \emph{external validity} lens to the attack evaluation domain: rather than assessing whether individual benchmark experiments are internally valid, we audit whether benchmarks collectively cover the attack surface they claim to evaluate.
\end{enumerate}

To address these issues, we construct a data-informed taxonomy from our set of papers, define a matrix for its attacks, and use that to audit current benchmark coverage. Our contributions are:

\begin{enumerate}[leftmargin=*,itemsep=2pt]
\item \textbf{A structured dataset of 2{,}521 unique attack groups} extracted from 932 arXiv papers (H1 2023--H1 2026), arranged into a 507-leaf taxonomy (430 attack types across five inference-time categories, 77 attack generation methods).

\item \textbf{A Target $\times$ Technique classification matrix.} We organize attacks into a 4$\times$6 matrix grounded in STRIDE~\citep{shostack2014threat}, where each attack tuple $(T, M)$ maps to a cell by adversarial target and bypassed defense layer, enabling systematic gap identification.

\item \textbf{A benchmark coverage audit.} We audit three major evaluation frameworks (HarmBench, InjecAgent, and AgentDojo) against the matrix, with three additional benchmarks (AdvBench, JailbreakBench, StrongREJECT) mapped in sensitivity analysis. The three primary benchmarks have zero overlapping cells and collectively cover at most 25\% of the matrix; the entire Service Disruption row and the Model Internals column have zero standardized evaluation across all six benchmarks.

\item \textbf{A set of quantitative characterizations of the field.} We discuss existing naming fragmentation, temporal evolution, and research concentrations which reveal structural properties of the field not visible at smaller-scale analysis.
\end{enumerate}

\noindent Together, these artifacts enable a shift from benchmark-internal validation---assessing whether a single benchmark measures what it claims---to benchmark-external validation: auditing whether the field's evaluation infrastructure collectively covers the threat surface it purports to address. This is the distinction \citet{liao2021we} identify as critical for evaluation maturity, applied here to the LLM security domain.

\section{Methodology}
\label{sec:method}

\subsection{Corpus and Extraction}
Our corpus of papers is comprised of 932 papers from arXiv, published between H1 2023 and H1 2026 (where H1/H2 denote January--June and July--December respectively), drawn from the Promptfoo LLM Security Database~\citep{PromptfooLMSecurityDB}. These papers cover adversarial attacks, jailbreaking, prompt injection, redteaming, and safety alignment of LLMs. The database indexes vulnerabilities across 803 LLM models, categorized by attack type (injection, jailbreak, extraction, denial-of-service, poisoning, prompt-leaking, side-channel), attack surface, and threat context; its scope is explicitly attack- and vulnerability-focused. While the database does not publish formal paper inclusion criteria, any resulting selection bias would strengthen our gap analysis, as over-representation of well-studied areas means tail categories are \emph{even more} understudied than we report (see \S\ref{sec:limitations}).

We used Gemini 3.1 Pro (\texttt{gemini-3.1-pro-preview}, temperature 0, structured JSON output mode) for structured extraction from each paper's text, following the LLM-driven literature processing paradigm validated at comparable scale by \citet{wu2025automatedreview}. The extraction prompt was scoped to inference-time attacks only, with explicit rejection criteria for training-phase techniques, generic category terms, and non-attack components. We pin the model version and provide all extraction prompts verbatim in Appendix~\ref{app:prompts}.

\subsection{Normalization and classification}

The raw extraction yielded 6{,}366 attack mentions across 863 papers (69 papers contained no inference-time attacks), which were then normalized to 2{,}521 unique attack groups through case normalization, alias merging, and deduplication (60.4\% reduction). The alias-merging step required over 50 manually curated variant clusters with explicit rules for blocking category-to-member conflations (e.g., preventing ``gradient-based'' from absorbing ``GCG''), resolving ambiguous abbreviations, and preserving genuinely distinct techniques that share surface names (full merge logic in released pipeline code). Each attack group was classified via a three-tier system:

\begin{itemize}[leftmargin=*,itemsep=1pt]
\item \textbf{Tier 1: Direct lookup} (2.1\%): name of extracted attack was an exact match against taxonomy leaf identifiers.
\item \textbf{Tier 2: Fuzzy matching} (2.7\%): name of extracted attack had string similarity via \texttt{SequenceMatcher} (cutoff 0.75).
\item \textbf{Tier 3: Semantic matching} (95.2\%): LLM-based classification against a compact leaf catalog with variant annotations.
\end{itemize}

\subsection{Validation} \label{sec:validation}

The final dataset yields \textbf{2{,}521 unique attack groups}. Three cross-reference checks identified 76 novelty conflicts or suspicious claims; after manual review, 33 false-novel claims were corrected and 1{,}133 of 2{,}521 groups retain verified novel status (an upper bound, as single-paper false claims without contradicting sources remain undetectable).

A human evaluation of 200 stratified Tier~3 classifications yielded \textbf{92.0\%} accuracy (95\% CI: [87.4\%, 95.0\%]), with only 2 of 16 errors crossing category boundaries (affecting matrix-level cell counts). Inter-annotator agreement on 115 double-coded samples yielded 86.1\% agreement at the matrix-cell level and leaf-level $\kappa = 0.70$. Re-classifying the same 200 attacks with Claude Opus 4.6 yielded 94.5\% target-row and 80.5\% matrix-cell agreement, with disagreements concentrated on technique-column boundaries (Appendix~\ref{app:validation}). An extraction recall spot-check found \textbf{100\%} precision (76/76) and \textbf{89.4\%} recall (76/85), with misses concentrated among lesser-known attacks~\citep{merullo2025on}. Full validation methodology, source attribution quality, and per-category breakdowns are in Appendix~\ref{app:validation} and the Datasheet (Appendix~\ref{app:datasheet}).

\section{Dataset}
\label{sec:dataset}

\subsection{Scope and taxonomy structure}
\label{sec:taxonomy}

In this paper, we focus on \emph{inference-time} attacks, defined as those executed through the model's input interface during deployment, requiring no access to the training pipeline or model weights~\citep{biggio2013evasion, papernot2017practical}. Training-phase attacks (data poisoning, backdoor injection, weight manipulation) are covered by existing surveys~\citep{xu2025survey} and require a fundamentally different threat model; we exclude them to ensure comprehensive coverage within our scope. The extraction pipeline also identified 414 training-phase attack types from the corpus; these are included in the released dataset but excluded from the taxonomy and matrix analysis below.

The taxonomy organizes the 2{,}521 unique attack groups into 507 leaves across five attack categories and a sixth method category (Attack Generation). Of these, 401 are populated by at least one attack group in the corpus; the remaining 106 represent attack types documented in the security literature or derivable from threat-model reasoning (e.g., \texttt{audio-steganography}, \texttt{calendar-meeting-invite-injection}) that no paper in the corpus mentioned, serving as placeholders for future coverage. These leaves were authored by the research team during taxonomy construction, not generated by the extraction pipeline. The taxonomy hierarchy was designed independently of the extraction pipeline: the LLM served as the extraction and classification instrument, while all structural decisions---including the 106 placeholder leaves---were made by the authors based on threat-model reasoning (construction process in Appendix~\ref{app:validation}). The resulting categories---Jailbreaking, Prompt Injection, Information Disclosure, Service Disruption, and Decoding Manipulation---align with prior surveys~\citep{xu2025survey, kim2026attacklandscape} while adding Service Disruption as a separate category because it targets availability rather than content. The 4$\times$6 matrix (Section~\ref{sec:matrix}) was chosen to be mutually exclusive and collectively exhaustive over the corpus; alternative decompositions are possible (see Limitations). Table~\ref{tab:category_stats} summarizes the distribution, while Figure~\ref{fig:taxonomy} (Appendix~\ref{app:taxonomy}) provides a visual representation.

  \begin{table}[t]                                 
  \caption{Attack distribution by category. Jailbreaking dominates the literature, while Service Disruption remains underexplored despite its high novelty rate (34.3\% vs.\ 11.0--31.4\% for other categories).}
  \label{tab:category_stats}
  \centering
  \small
  \begin{tabular}{lrrrrrr}
  \toprule
  Category & Leaves & Mentions & Unique & Papers & Novel & Novelty \% \\
  \midrule
  Jailbreaking        & 316 & 2{,}869 & 1{,}350 & 712 &    628 & 21.9 \\
  Prompt Injection     &  47 &    472 &    267 & 192 &    148 & 31.4 \\
  Info.\ Disclosure    &  35 &    101 &     84 &  55 &     40 & 39.6 \\
  Decoding Manip.      &   6 &     82 &     62 &  44 &     34 & 41.5 \\
  Service Disruption   &  26 &    134 &    103 &  49 &     46 & 34.3 \\
  \midrule
  \textbf{Subtotal}   & \textbf{430} & \textbf{3{,}658} & --- & --- & \textbf{896} & \textbf{24.5} \\
  \addlinespace
  \textit{Attack Generation\textsuperscript{$\dagger$}} & 77 & 2{,}465 & 492 & 606 & 270 & 11.0 \\
  \bottomrule
  \end{tabular}   
  \vspace{2pt}
  \par\noindent{\footnotesize \textsuperscript{$\dagger$}Orthogonal category describing \emph{how} attacks are generated, not attack types.}
  \end{table}

Each category is defined by a distinct adversarial objective and defensive requirement; full descriptions with subcategory breakdowns are in Appendix~\ref{app:categories}. The five attack categories are: Jailbreaking (316 leaves; bypassing safety alignment~\citep{zou2023universal}), Prompt Injection (47; via external data sources~\citep{greshake2023not}), Information Disclosure (35; extracting non-public data~\citep{carlini2021extracting}), Service Disruption (26; degrading availability~\citep{feiglin2026benchoverflow}), and Decoding Manipulation (6; intervening in generation, partially requiring white-box access; see Appendix~\ref{app:categories}). A sixth \emph{Attack Generation} category (77 leaves) captures methods for \emph{discovering} attacks (e.g., GCG~\citep{zou2023universal}, PAIR~\citep{chao2023pair}, AutoDAN~\citep{liu2024autodan}) rather than attack types themselves; these are classified in the matrix by their \emph{output} payload's mechanism rather than their search method (see worked example in Section~\ref{sec:matrix}).
\subsection{Target \texorpdfstring{$\times$}{x} Technique matrix}
\label{sec:matrix}

The taxonomy and the matrix serve distinct purposes. The \emph{taxonomy} (Table~\ref{tab:category_stats}) organizes attacks by their \emph{literature role} (how they are studied and published) and provides a browsable hierarchy for the released dataset. The \emph{matrix} (Table~\ref{tab:matrix}) provides an orthogonal classification by \emph{operational role} (what target the adversary pursues and which mechanism is used) and is the basis for the benchmark coverage audit. Because these axes are independent, the per-category counts in Table~\ref{tab:category_stats} and the per-cell counts in Table~\ref{tab:matrix} differ.\footnote{The taxonomy and matrix axes are independent: attacks counted under Jailbreaking in Table~\ref{tab:category_stats} may appear under Service Disruption in Table~\ref{tab:matrix}, and vice versa.}

\paragraph{Worked example.} GCG~\citep{zou2023universal} appears under \emph{Attack Generation} in the taxonomy (gradient-based optimization) but maps to Safety Bypass $\times$ Obfuscation in the matrix (its output [adversarial token suffixes] bypasses input filters). References to ``GCG'' (the method) and ``adversarial suffix'' (its product) are tracked as separate attack groups, preventing double-counting.

\paragraph{Target rows (STRIDE grounding).} The four target rows are grounded in the STRIDE threat model~\citep{shostack2014threat}: Safety \& Alignment Bypass (Tampering), System \& Tool Hijacking (Elevation of Privilege), Information Exfiltration (Information Disclosure), and Service Disruption (Denial of Service). Spoofing and Repudiation do not apply to inference-time attacks; training-phase targets are excluded.

An explicit \emph{output-vs-execution rule} resolves the boundary between the first two targets: Safety Bypass for attacks that succeed via harmful \emph{output}, System Hijacking for those requiring model \emph{action} (Appendix~\ref{app:taxonomy}).

\paragraph{Technique columns (bypass-target principle).} Each column is defined by which \emph{defense layer} the attack circumvents, an organizing principle that emerged from the taxonomy's natural mechanism clusters: Instructional (instruction priority), Persuasion \& Deception (alignment/RLHF), Obfuscation (input filtering), Cross-Modal (text-only safety), Indirect Injection (trust boundaries), and Model Internals (architecture/representations). Formal defense-layer definitions are in Appendix~\ref{app:taxonomy}.

\paragraph{Classification procedure.} Of the 507 leaves, 333 receive their technique column from the taxonomy structure directly; the remaining 174 were individually assigned by defense layer bypassed (details in Appendix~\ref{app:taxonomy}).

\subsection{Released artifacts}
\label{sec:release}

We release six artifacts on HuggingFace with Croissant metadata: (1)~the 507-leaf attack taxonomy as hierarchical JSON with Target~$\times$~Technique classifications, with an interactive browser for local exploration; (2)~enriched attack records (6{,}366 entries) with source paper, match tier, and novelty status; (3)~pre-computed matrix statistics and temporal heatmaps; (4)~the extraction and classification pipeline code; (5)~a source paper manifest (932 entries); and (6)~a codebook explaining the manual verification process for the three stage output. All data derives from publicly available arXiv papers. The taxonomy and attack records are released under CC-BY-4.0; the pipeline code under MIT license.

\section{Analysis and applications}
\label{sec:analysis}

We demonstrate the dataset's utility through three separate analyses that reveal structural properties of the field invisible to individual-paper review, followed by a benchmark coverage audit that identifies systematic evaluation gaps.

\subsection{Target \texorpdfstring{$\times$}{x} Technique distribution}

Table~\ref{tab:matrix} presents the Target $\times$ Technique matrix with taxonomy leaf counts per cell. The distribution reveals extreme concentration: Safety \& Alignment Bypass accounts for 74.4\% of all taxonomy leaves (377/507), consistent with the field's broader focus on jailbreaking~\citep{xu2025survey,wei2024jailbroken}. Obfuscation and Persuasion are the two largest technique columns (132 and 122 leaves respectively), reflecting the diversity of encoding schemes and social-engineering framings that researchers have documented. For example, Persuasion applied to System Hijacking is realized by attacks like \texttt{action-hijacking}, which deceives the model into misusing legitimate tools through deceptive framing. Of the five empty cells, three are corpus gaps ($\cdot$) and two require compound attacks that bypass two defense layers simultaneously ($\diamond$); see Appendix~\ref{app:taxonomy}.

\begin{table}[t]
\caption{Target $\times$ Technique matrix: taxonomy leaf counts per cell. Cells marked ``$\bullet$'' are corpus gaps (structurally coherent but absent from the literature); cells marked ``$\diamond$'' require compound attacks bypassing two defense layers (see Appendix~\ref{app:taxonomy}).}
\label{tab:matrix}
\centering
\small
\setlength{\tabcolsep}{4pt}
\begin{tabular}{l rrrrrr r}
\toprule
& \rotatebox{55}{Instruct.} & \rotatebox{55}{Persuasion} & \rotatebox{55}{Obfusc.} & \rotatebox{55}{Cross-Mod.} & \rotatebox{55}{Indirect Inj.} & \rotatebox{55}{Model Int.}
& \textbf{Total} \\
\midrule
Safety \& Align.\ Bypass & 78 & 108 & 125 & 53 & $\diamond$ & 13 & \textbf{377} \\
System \& Tool Hijacking & 9 & 4 & 2 & 8 & 45 & $\bullet$ & \textbf{68} \\
Info.\ Exfiltration & 10 & 8 & 4 & $\diamond$ & $\bullet$ & 13 & \textbf{35} \\
Service Disruption & 13 & 2 & 1 & $\bullet$ & 4 & 7 & \textbf{27} \\
\midrule
\textbf{Total} & \textbf{110} & \textbf{122} & \textbf{132} & \textbf{61} & \textbf{49} & \textbf{33} & \textbf{507} \\
\bottomrule
\end{tabular}
\end{table}

The matrix reveals three patterns:

\begin{enumerate}[leftmargin=*,itemsep=2pt]
\item \textbf{Indirect Injection is almost exclusive to System Hijacking.} 45 of 49 Indirect Injection leaves target System \& Tool Hijacking, with the remaining 4 targeting Service Disruption (e.g., triggering costly tool calls). This reflects a structural reality: indirect injection operates by embedding instructions in external data, which is primarily effective when the model has tool-calling or agentic capabilities.

\item \textbf{Obfuscation, Persuasion, and Cross-Modal concentrate in Safety Bypass.} These three technique columns place 87--95\% of their leaves in the Safety Bypass row (125/132, 108/122, and 53/61 respectively). Bypassing alignment is the dominant use case for encoding tricks, social engineering, and multimodal attacks; other targets are pursued through Instructional or Indirect Injection mechanisms.

\item \textbf{Instructional techniques are the most broadly distributed.} Instructional attacks appear in all four target rows with substantial counts (78, 9, 10, 13), reflecting that instruction manipulation is the most versatile attack primitive, applicable whether the goal is harmful content, unauthorized execution, data extraction, or resource exhaustion.
\end{enumerate}

\noindent All three patterns are robust to model choice (Appendix~\ref{app:validation}).

\subsection{Naming fragmentation}

Naming fragmentation is severe. The GCG attack, defined in ~\citet{zou2023universal} appears under 29 distinct surface forms across 376 papers, making it the single most referenced yet most inconsistently named attack. Similarly, AutoDAN, defined in~\citet{liu2024autodan}, appears under 11 distinct forms across 259 papers. At the taxonomy level, 60 leaf nodes function as ``catch-alls,'' each absorbing 10+ distinct attack techniques. The top catch-all (\texttt{adversarial-perturbation-on-images}) maps 126 unique attacks from the pipeline (see Appendix~\ref{app:catchall} for recommended splits). This fragmentation inflates apparent attack counts, complicates cross-paper comparison, and impedes reproducible evaluation: motivating the standardized naming our dataset provides.

\subsection{Temporal evolution}
\begin{figure}[t]
\centering
\includegraphics[width=\linewidth]{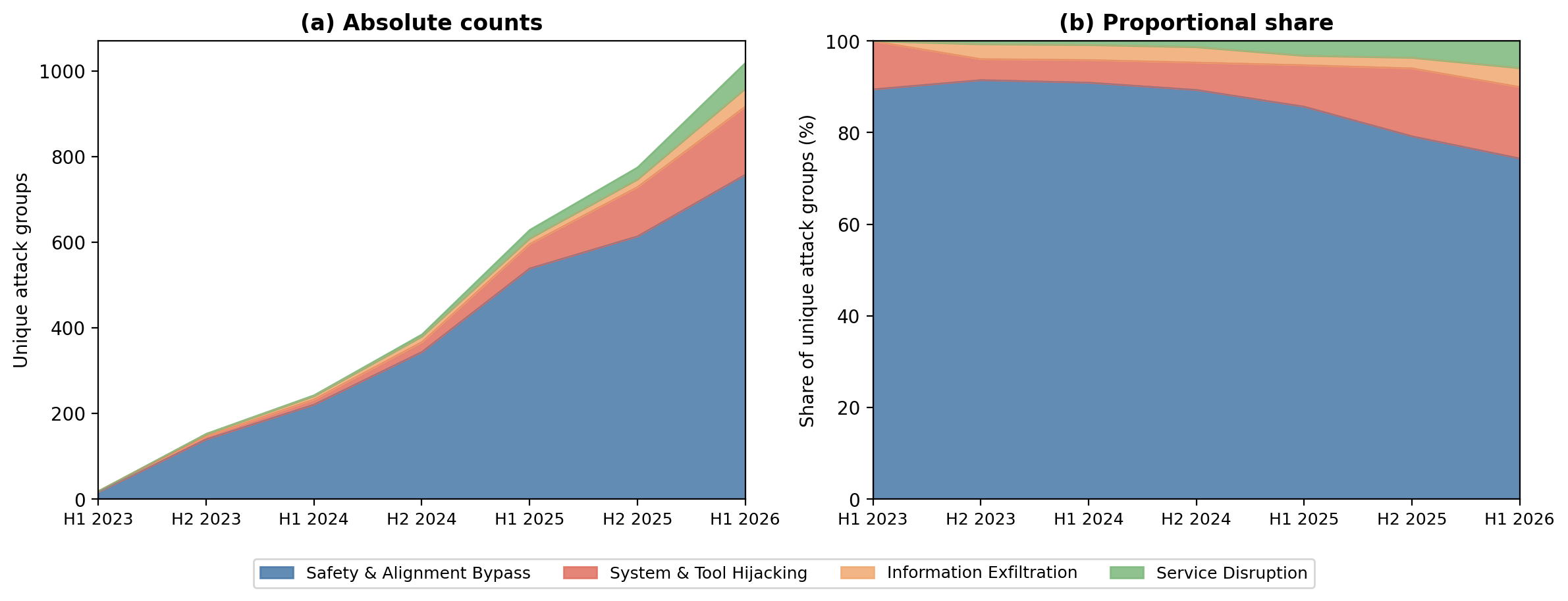}
\caption{Unique attack groups by target category over time. (a)~Absolute counts show broad-based growth across all four target categories. (b)~Proportional share reveals a gradual diversification: Safety Bypass declined from 92\% to 74\% of unique attack groups between H2 2023 and H1 2026, while System \& Tool Hijacking grew from 5\% to 16\%. Target-row assignments are pipeline-dependent (94.5\% cross-model agreement; see Appendix~\ref{app:validation}).}
\label{fig:temporal_targets}
\end{figure}

The field shows a clear temporal structure (Table~\ref{tab:corpus}, Appendix~\ref{app:corpus}). The novelty rate spiked at 26.9\% in H2 2023 before stabilizing at ${\sim}$18\%, indicating that approximately one-fifth of newly reported attacks genuinely extend the taxonomy. This figure should be treated as an apparent rate due to potential confounds with pipeline recall (see Appendix~\ref{app:validation} for a detailed analysis). Safety \& Alignment Bypass has remained the dominant focus (83.8\% of mapped mentions), though its share has declined from 92\% to 74\% as agentic categories (System Hijacking, Information Exfiltration) have grown (Figure~\ref{fig:temporal_targets}). This diversification suggests the benchmark coverage gap may be \emph{widening}, as no benchmark covers these emerging categories.

\subsection{Benchmark coverage audit}
\label{sec:benchmark}

The Target $\times$ Technique matrix enables a systematic audit of evaluation coverage. We mapped three major LLM attack benchmarks onto the matrix by matching each evaluated method to its taxonomy leaf and corresponding matrix cell (procedure details in Appendix~\ref{app:benchmark_mapping}). The three benchmarks are: HarmBench~\citep{mazeika2024harmbench} (18 red-teaming methods, 510 behaviors), InjecAgent~\citep{zhan2024injecagent} (indirect prompt injection, 1{,}054 cases), and AgentDojo~\citep{debenedetti2024agentdojo} (dynamic agent environments, 629 test cases).

\begin{table}[t]
\caption{Benchmark coverage matrix. H = HarmBench, I = InjecAgent, A = AgentDojo. Superscript $\circ$ = partial coverage. Combined coverage: at most 6 of 24 cells (25\%); sensitivity analysis in Appendix~\ref{app:benchmark_mapping}.}
\label{tab:benchmark}
\centering
\small
\begin{tabular}{l cccccc}
\toprule
& Instruct. & Persuasion & Obfusc. & Cross-Mod. & Indirect Inj. & Model Int. \\
\midrule
Safety \& Align.\ Bypass & H$^\circ$ & H & H & H$^\circ$ & --- & --- \\
System \& Tool Hijacking & --- & --- & --- & --- & I, A & --- \\
Info.\ Exfiltration & --- & --- & --- & --- & I, A & --- \\
Service Disruption & --- & --- & --- & --- & --- & --- \\
\bottomrule
\end{tabular}
\end{table}

\paragraph{Per-benchmark summary.} HarmBench's 18 red-teaming methods span four technique columns (Obfuscation, Persuasion, Instructional, Cross-Modal) but target exclusively Safety \& Alignment Bypass. InjecAgent (1{,}054 cases) and AgentDojo (629 test cases) both cover Indirect Injection for System \& Tool Hijacking and Information Exfiltration. The three benchmarks have \textit{zero overlapping cells}: HarmBench and the agent benchmarks test completely disjoint parts of the attack surface, and no benchmark cross-validates another's coverage area.

\paragraph{Critical blind spots.} Among the six publicly available benchmarks audited, at least 18 of 24 cells lack any standardized evaluation. Proprietary or custom red-teaming suites may cover additional cells; our audit therefore establishes a \emph{lower bound} on the public evaluation gap. The uncovered cells include:

\begin{itemize}[leftmargin=*,itemsep=1pt]
\item \textbf{Service Disruption} (entire row): resource exhaustion, logic bombs, ReDoS, denial-of-wallet. No benchmark covers any technique column for this target.
\item \textbf{Model Internals} (entire column): activation steering, embedding inversion, membership inference, side-channel probing, refusal-subspace ablation. No benchmark evaluates these mechanisms, yet on identical AdvBench behaviors where GCG (benchmarked) achieves 69\% ASR, logit suppression~\citep{li2024logitsuppression} (unbenchmarked) achieves 96\%, but through a mechanism no benchmark evaluates.
\item \textbf{Safety Bypass $\times$ Indirect Injection}: no benchmark tests whether indirect injection via external data can bypass safety alignment to produce harmful content, as opposed to hijacking agent actions.
\end{itemize}

In the terms of~\citet{liao2021we}'s external-validity framework, these gaps reflect both \emph{dataset misalignment} and deeper \emph{metrics misalignment} (see Appendix~\ref{app:benchmark_mapping}). That three independently designed benchmarks converge on the same narrow slice suggests designers gravitate toward the most visible attack modalities, creating systematic blind spots where attacks are most costly in production. Three additional benchmarks (AdvBench, JailbreakBench, StrongREJECT) add no new cells (sensitivity analysis in Appendix~\ref{app:benchmark_mapping}).

\paragraph{Robustness.} The coverage gap is stable under perturbations: re-classifying 200 attacks with Claude Opus 4.6 yields 94.5\% target-row agreement and preserves all three structural patterns (Appendix~\ref{app:validation}); excluding partial-coverage cells reduces coverage from 6 to 4 of 24, still $\leq$25\%. An alternative leaf-weighted metric (${\sim}$80\%) is higher because covered cells coincide with the most-studied region, but conflates research attention with threat importance; the structural gap operates at the level of STRIDE threat categories and is invariant to matrix resolution (Appendix~\ref{app:benchmark_mapping}).

\section{Discussion}
\label{sec:discussion}

\paragraph{Research concentration vs.\ threat distribution.} Research attention and threat coverage are mismatched. Safety \& Alignment Bypass dominates the literature (84\% of mentions), yet real-world deployments face threats across all four inference-time target categories. Service Disruption (resource exhaustion, denial-of-wallet) is arguably extremely high-stakes in production settings, as a DoS attack on an LLM can impose unbounded cost, however it receives minimal research attention and zero public benchmark coverage. Moreover, Service Disruption, Information Disclosure, and Decoding Manipulation have the highest novelty rates (34.3\%, 39.6\%, and 41.5\% respectively), indicating rapid expansion of precisely the areas benchmarks do not cover---in contrast to Attack Generation Methods (11.0\%), where the focus has shifted from discovery to refinement. These gaps are empirically validated: published attacks in uncovered cells achieve 46$\times$ token amplification~\citep{kumar2026overthinkslowdownattacksreasoning} and 100$\times$ latency increases~\citep{li2025thinktrapdenialofserviceattacksblackbox} (Service Disruption), or completely disable safety training via single weight-space interventions (Model Internals), yet lack any standardized evaluation. Closing these gaps requires new measurement infrastructure that existing prompt-level benchmarks cannot provide. A Service Disruption benchmark would need: (i)~per-request resource accounting (tokens generated, wall-clock latency, API cost) alongside correctness metrics; (ii)~a threat model specifying attacker budget (e.g., queries per minute) and defender capacity; and (iii)~attack scenarios spanning computational exhaustion (e.g., reasoning-loop amplification), functional degradation (e.g., irrelevant-distraction injection), and denial-of-wallet (e.g., recursive tool calls)---the three subcategories our taxonomy identifies.
\paragraph{Using the matrix for benchmark design.} The primary contribution of this work is not the coverage finding itself---that benchmarks concentrate on safety-bypass jailbreaking is directionally expected---but the reusable framework that makes coverage gaps \emph{measurable, comparable, and actionable}. The released matrix and taxonomy provide a systematic workflow for benchmark designers: identify the target cells within the benchmark's intended scope, select taxonomy leaves within those cells as candidate attack methods, design test cases grounded in the leaf descriptions and cited papers, and validate coverage against the full matrix to document which cells remain outside scope. This process transforms benchmark design from ad hoc attack selection into an auditable gap analysis, and provides a common coordinate system for comparing what different benchmarks cover. Appendix~\ref{app:benchmark_design} walks through this workflow end-to-end for the Service Disruption $\times$ Instructional cell, assembling published evidence from four corpus papers into a concrete benchmark specification. As new benchmarks emerge, they can be mapped onto the same matrix, enabling the community to track whether coverage gaps are closing over time without requiring a new survey for each assessment.

\paragraph{Naming fragmentation impedes progress.} The 29 surface forms for GCG across 376 papers illustrate that the field lacks consensus on attack naming. This is not merely an aesthetic issue: it inflates apparent novelty, complicates meta-analysis, and makes it impossible to determine from paper titles alone whether two works study the same technique. Our standardized taxonomy IDs and cross-paper deduplication provide a foundation for addressing this problem.

\section{Limitations}
\label{sec:limitations}

\paragraph{Pipeline.} The pipeline relies on Gemini 3.1 Pro for both extraction and classification, so a systematic blind spot for a category would be invisible. The 89.4\% extraction recall (\S\ref{sec:validation}) is familiarity-dependent: misses are distributed across the taxonomy rather than systematically absent from particular branches. The multi-model \emph{classification} comparison (94.5\% target-row agreement between Gemini and Claude; Appendix~\ref{app:validation}) mitigates this risk for the classification stage, while the human recall spot-check mitigates it for extraction. The benchmark audit (\S\ref{sec:benchmark}) is unaffected, as benchmark methods are mapped by manually reading documentation. Additionally, the frequency-dependent extraction gap~\citep{merullo2025on} implies that novel or less-established attacks may be undercounted, meaning the observed novelty stabilization at ${\sim}$18\% (\S\ref{sec:analysis}) should be treated as an apparent rate rather than a precise measurement of field maturation.

\paragraph{Scope.} The corpus is limited to English arXiv papers from the Promptfoo LLM Security Database~\citep{PromptfooLMSecurityDB}, which does not publish formal paper collection criteria (see \S\ref{sec:method}). To estimate representativeness, we checked the reference list of an independent survey~\citep{kim2026attacklandscape} against the corpus: of 102 cited arXiv papers, 76 fall outside the database's documented scope (defenses, surveys, foundational models, infrastructure); of the 26 in-scope attack papers, 8 (30.8\%) appear in the corpus. However, the attack \emph{techniques} described by all 18 missing papers are represented in the taxonomy through citing papers already in the corpus, indicating that source-paper gaps do not produce taxonomy gaps. Any remaining source-specific over-representation would mean tail categories are \emph{more} understudied than we report, strengthening the gap analysis. We audited six public benchmarks; proprietary or custom red-teaming suites may cover additional cells, so our audit establishes a lower bound on coverage gaps. The corpus is weighted toward recent publications (305 papers in H1 2026 vs.\ 3 in H1 2023), which may over-represent current trends relative to foundational work. The benchmark coverage audit is unaffected by this temporal skew, as benchmarks are mapped to matrix cells via manual documentation review rather than corpus statistics; the uncovered cells (Service Disruption, Model Internals) remain empty regardless of which corpus period is considered.

\section{Related Work}
\label{sec:related}

Several recent works organize the landscape of adversarial attacks on language models, each from a distinct vantage point. \citet{xu2025survey} provide a narrative survey partitioned by lifecycle phase (training, inference, deployment), covering roughly 150 papers and ${\sim}$40 named attacks; however, their breadth across the full lifecycle comes at the cost of granular depth within any single phase. \citet{kim2026attacklandscape} present a systematization of knowledge for agentic AI, crossing six attack vectors with CIA-grounded security risks across 128 papers. While authoritative for autonomous systems, they scope their taxonomy to agent-specific threats such as tool poisoning and memory manipulation, largely excluding the direct-to-model inference attacks that affect non-agentic LLMs. Notably, both surveys' coverage maps onto the same concentrated matrix region as the benchmarks we audit: Xu et al.'s inference-time attacks fall within Safety \& Alignment Bypass, and Kim et al.'s within System Hijacking $\times$ Indirect Injection, while neither survey systematically addresses Service Disruption or Model Internals.

Focused studies provide higher resolution but narrower scope. \citet{liu2024formalizing} formalize prompt injection with a mathematical framework and benchmark five strategies against ten defenses, offering rigorous depth on a single attack family but omitting the broader adversarial landscape (e.g., gradient-based perturbations, decoding-time manipulations, or service disruption). \citet{wang2026sok} systematize jailbreak \emph{defenses} along six dimensions---from intervention stage to technical paradigm---offering a complementary defense-side view but no attack taxonomy. \citet{grosse2024practical} survey 271 industrial practitioners to measure the divergence between academic threat models and real-world AI usage, finding that research assumptions are often too generous about attacker access; their work serves as a valuable socio-technical critique but does not construct an attack taxonomy.

Our work differs in three respects. First, we construct the taxonomy \emph{data-informed} from 932 papers---6--18$\times$ larger than comparable corpora---using automated extraction rather than manual curation, yielding 401 populated leaves across 19 of 24 matrix cells compared to Xu et al.'s ${\sim}$40 named attacks across the full lifecycle and Kim et al.'s focus on one matrix region (System Hijacking $\times$ Indirect Injection). Second, we decouple the browsable hierarchy (taxonomy) from an operational risk view (the 4$\times$6 Target $\times$ Technique matrix). Third, we leverage this matrix for a benchmark coverage audit, revealing at most 25--30\% cell coverage with zero overlap between benchmarks, a structural gap invisible to individual-benchmark evaluation.

\section{Conclusion}
\label{sec:conclusion}
The central finding of this work is structural: three independently designed benchmarks converge on the same narrow slice of the attack surface with zero overlap, leaving entire STRIDE threat categories without any standardized public evaluation. We establish this through a 507-leaf taxonomy extracted from 932 papers, organized into both a browsable hierarchy and a Target $\times$ Technique matrix grounded in STRIDE~\citep{shostack2014threat} and the bypass-target principle. The Service Disruption row (27 leaves) and Model Internals column (33 leaves) receive zero coverage across all six benchmarks audited. The taxonomy, enriched attack records, and benchmark coverage mappings are released to support three concrete next steps: (1)~developing benchmarks for uncovered matrix cells, beginning with Service Disruption, which has the highest novelty rate yet zero evaluation coverage; (2)~adopting standardized attack identifiers to reduce the naming fragmentation that currently obscures cross-paper comparison; and (3)~extending the corpus to non-English and non-arXiv sources to test whether the concentration patterns we observe reflect genuine research gaps or artifacts of corpus scope.

\bibliographystyle{plainnat}
\bibliography{references_v3}

@book{shostack2014threat,
  title={Threat Modeling: Designing for Security},
  author={Shostack, Adam},
  year={2014},
  publisher={John Wiley \& Sons}
}

@misc{zou2023universal,
      title={Universal and Transferable Adversarial Attacks on Aligned Language Models}, 
      author={Andy Zou and Zifan Wang and Nicholas Carlini and Milad Nasr and J. Zico Kolter and Matt Fredrikson},
      year={2023},
      eprint={2307.15043},
      archivePrefix={arXiv},
      primaryClass={cs.CL}
}

@inproceedings{wei2024jailbroken,
    author = {Wei, Alexander and Haghtalab, Nika and Steinhardt, Jacob},
    title = {Jailbroken: how does LLM safety training fail?},
    year = {2023},
    publisher = {Curran Associates Inc.},
    address = {Red Hook, NY, USA},
    booktitle = {Proceedings of the 37th International Conference on Neural Information Processing Systems},
    articleno = {3508},
    numpages = {32},
    location = {New Orleans, LA, USA},
    series = {NIPS '23}
}

@inproceedings{qi2023finetuning,
     author = {Qi, Xiangyu and Zeng, Yi and Xie, Tinghao and Chen, Pin-Yu and Jia, Ruoxi and Mittal, Prateek and Henderson, Peter},
     booktitle = {International Conference on Learning Representations},
     editor = {B. Kim and Y. Yue and S. Chaudhuri and K. Fragkiadaki and M. Khan and Y. Sun},
     pages = {30988--31043},
     title = {Fine-tuning Aligned Language Models Compromises Safety, Even When Users Do Not Intend To!},
     volume = {2024},
     year = {2024}
}

@misc{xu2025survey,
      title={A Survey of Attacks on Large Language Models}, 
      author={Wenrui Xu and Keshab K. Parhi},
      year={2025},
      eprint={2505.12567},
      archivePrefix={arXiv},
      primaryClass={cs.CR}
}

@article{kim2026attacklandscape,
    title={The Attack and Defense Landscape of Agentic {AI}: A Comprehensive Survey},
    author={Kim, Juhee and Liu, Xiaoyuan and Wang, Zhun and Qiu, Shi and Li, Bo and Guo, Wenbo and Song, Dawn},
    journal={arXiv preprint arXiv:2603.11088},
    year={2026},
    note={Accepted to USENIX Security 2026}
  }

@inproceedings{wang2026sok,
    title={SoK: Evaluating Jailbreak Guardrails for Large Language Models},
    author={Wang, Xunguang and Ji, Zhenlan and Wang, Wenxuan and Li, Zongjie and Wu, Daoyuan and Wang, Shuai},
    booktitle={IEEE Symposium on Security and Privacy (SP)},
    year={2026}
}

@inproceedings{liu2024formalizing,
    author = {Liu, Yupei and Jia, Yuqi and Geng, Runpeng and Jia, Jinyuan and Gong, Neil Zhenqiang},
    title = {Formalizing and benchmarking prompt injection attacks and defenses},
    year = {2024},
    isbn = {978-1-939133-44-1},
    publisher = {USENIX Association},
    address = {USA},
    booktitle = {Proceedings of the 33rd USENIX Conference on Security Symposium},
    articleno = {103},
    numpages = {17},
    location = {Philadelphia, PA, USA},
    series = {SEC '24}
}

@inproceedings{grosse2024practical,
    author = {Grosse, Kathrin and Bieringer, Lukas and Besold, Tarek R. and Alahi, Alexandre},
    title = {Towards more practical threat models in artificial intelligence security},
    year = {2024},
    isbn = {978-1-939133-44-1},
    publisher = {USENIX Association},
    address = {USA},
    booktitle = {Proceedings of the 33rd USENIX Conference on Security Symposium},
    articleno = {274},
    numpages = {18},
    location = {Philadelphia, PA, USA},
    series = {SEC '24}
}

@inproceedings{liao2021we,
    title={Are We Learning Yet? A Meta Review of Evaluation Failures Across Machine Learning},
    author={Thomas Liao and Rohan Taori and Inioluwa Deborah Raji and Ludwig Schmidt},
    booktitle={Thirty-fifth Conference on Neural Information Processing Systems Datasets and Benchmarks Track (Round 2)},
    year={2021}
}

@inproceedings{mazeika2024harmbench,
    author = {Mazeika, Mantas and Phan, Long and Yin, Xuwang and Zou, Andy and Wang, Zifan and Mu, Norman and Sakhaee, Elham and Li, Nathaniel and Basart, Steven and Li, Bo and Forsyth, David and Hendrycks, Dan},
    title = {HarmBench: a standardized evaluation framework for automated red teaming and robust refusal},
    year = {2024},
    publisher = {JMLR.org},
    booktitle = {Proceedings of the 41st International Conference on Machine Learning},
    articleno = {1431},
    numpages = {44},
    location = {Vienna, Austria},
    series = {ICML'24}
}

@inproceedings{zhan2024injecagent,
    title = "{I}njec{A}gent: Benchmarking Indirect Prompt Injections in Tool-Integrated Large Language Model Agents",
    author = "Zhan, Qiusi and Liang, Zhixiang and Ying, Zifan and Kang, Daniel",
    editor = "Ku, Lun-Wei and Martins, Andre and Srikumar, Vivek",
    booktitle = "Findings of the Association for Computational Linguistics: ACL 2024",
    month = aug,
    year = "2024",
    address = "Bangkok, Thailand",
    publisher = "Association for Computational Linguistics",
    doi = "10.18653/v1/2024.findings-acl.624",
    pages = "10471--10506"
}

@inproceedings{debenedetti2024agentdojo,
    title={AgentDojo: A Dynamic Environment to Evaluate Prompt Injection Attacks and Defenses for {LLM} Agents},
    author={Edoardo Debenedetti and Jie Zhang and Mislav Balunovic and Luca Beurer-Kellner and Marc Fischer and Florian Tram{\`e}r},
    booktitle={The Thirty-eighth Conference on Neural Information Processing Systems Datasets and Benchmarks Track},
    year={2024}
}

@inproceedings{chao2023pair,
  author={Chao, Patrick and Robey, Alexander and Dobriban, Edgar and Hassani, Hamed and Pappas, George J. and Wong, Eric},
  booktitle={2025 IEEE Conference on Secure and Trustworthy Machine Learning (SaTML)}, 
  title={Jailbreaking Black Box Large Language Models in Twenty Queries}, 
  year={2025},
  volume={},
  number={},
  pages={23-42},
  doi={10.1109/SaTML64287.2025.00010}}

@inproceedings{liu2024autodan,
    title={Auto{DAN}: Generating Stealthy Jailbreak Prompts on Aligned Large Language Models},
    author={Xiaogeng Liu and Nan Xu and Muhao Chen and Chaowei Xiao},
    booktitle={The Twelfth International Conference on Learning Representations},
    year={2024}
}

@inproceedings{li2023deepinception,
    title={DeepInception: Hypnotize Large Language Model to Be Jailbreaker},
    author={Xuan Li and Zhanke Zhou and Jianing Zhu and Jiangchao Yao and Tongliang Liu and Bo Han},
    booktitle={NeurIPS Safe Generative AI Workshop 2024},
    year={2024}
}

@inproceedings{mehrotra2024tree,
    author = {Mehrotra, Anay and Zampetakis, Manolis and Kassianik, Paul and Nelson, Blaine and Anderson, Hyrum and Singer, Yaron and Karbasi, Amin},
    title = {Tree of attacks: jailbreaking black-box LLMs automatically},
    year = {2024},
    isbn = {9798331314385},
    publisher = {Curran Associates Inc.},
    address = {Red Hook, NY, USA},
    booktitle = {Proceedings of the 38th International Conference on Neural Information Processing Systems},
    articleno = {1952},
    numpages = {41},
    location = {Vancouver, BC, Canada},
    series = {NIPS '24}
}

@inproceedings{yu2024gptfuzzer,
    author = {Jiahao Yu and Xingwei Lin and Zheng Yu and Xinyu Xing},
    title = {{LLM-Fuzzer}: Scaling Assessment of Large Language Model Jailbreaks},
    booktitle = {33rd USENIX Security Symposium (USENIX Security 24)},
    year = {2024},
    isbn = {978-1-939133-44-1},
    address = {Philadelphia, PA},
    pages = {4657--4674},
    publisher = {USENIX Association},
    month = aug
}

@inproceedings{russinovich2024crescendo,
    author = {Russinovich, Mark and Salem, Ahmed and Eldan, Ronen},
    title = {Great, now write an article about that: the crescendo multi-turn LLM jailbreak attack},
    year = {2025},
    isbn = {978-1-939133-52-6},
    publisher = {USENIX Association},
    address = {USA},
    booktitle = {Proceedings of the 34th USENIX Conference on Security Symposium},
    articleno = {125},
    numpages = {20},
    location = {Seattle, WA, USA},
    series = {SEC '25}
}

@inproceedings{wolf2023fundamental,
    author = {Wolf, Yotam and Wies, Noam and Avnery, Oshri and Levine, Yoav and Shashua, Amnon},
    title = {Fundamental limitations of alignment in large language models},
    year = {2024},
    publisher = {JMLR.org},
    booktitle = {Proceedings of the 41st International Conference on Machine Learning},
    articleno = {2176},
    numpages = {34},
    location = {Vienna, Austria},
    series = {ICML'24}
}

@article{kostikova2025lllms,
    author = {Kostikova, Aida and Wang, Zhipin and Bajri, Deidamea and P\"{u}tz, Ole and Paa\ss{}en, Benjamin and Eger, Steffen},
    title = {LLLMs: A Data-Driven Survey of Evolving Research on Limitations of Large Language Models},
    year = {2026},
    issue_date = {August 2026},
    publisher = {Association for Computing Machinery},
    address = {New York, NY, USA},
    volume = {58},
    number = {11},
    issn = {0360-0300},
    doi = {10.1145/3801096},
    journal = {ACM Comput. Surv.},
    month = apr,
    articleno = {282},
    numpages = {33}
}

@misc{PromptfooLMSecurityDB,
  author = {{promptfoo}},
  title = {LM Security Database},
  year = {2026},
  url = {https://www.promptfoo.dev/lm-security-db/},
  note = {Accessed: 4/10/26}
}

@inproceedings{biggio2013evasion,
    author = {Biggio, Battista and Corona, Igino and Maiorca, Davide and Nelson, Blaine and \v{S}rndi\'{c}, Nedim and Laskov, Pavel and Giacinto, Giorgio and Roli, Fabio},
    title = {Evasion attacks against machine learning at test time},
    year = {2013},
    isbn = {9783642409936},
    publisher = {Springer-Verlag},
    address = {Berlin, Heidelberg},
    doi = {10.1007/978-3-642-40994-3_25},
    booktitle = {Machine Learning and Knowledge Discovery in Databases: European Conference, {ECML PKDD} 2013, Proceedings, Part {III}},
    pages = {387–402},
    numpages = {16},
    location = {Prague, Czech Republic},
    series = {ECMLPKDD'13}
}

@inproceedings{papernot2017practical,
    author = {Papernot, Nicolas and McDaniel, Patrick and Goodfellow, Ian and Jha, Somesh and Celik, Z. Berkay and Swami, Ananthram},
    title = {Practical Black-Box Attacks against Machine Learning},
    year = {2017},
    isbn = {9781450349444},
    publisher = {Association for Computing Machinery},
    address = {New York, NY, USA},
    doi = {10.1145/3052973.3053009},
    booktitle = {Proceedings of the 2017 ACM on Asia Conference on Computer and Communications Security},
    pages = {506–519},
    numpages = {14},
    location = {Abu Dhabi, United Arab Emirates},
    series = {ASIA CCS '17}
}

@inproceedings{gong2023figstep,
    author = {Gong, Yichen and Ran, Delong and Liu, Jinyuan and Wang, Conglei and Cong, Tianshuo and Wang, Anyu and Duan, Sisi and Wang, Xiaoyun},
    title = {FigStep: jailbreaking large vision-language models via typographic visual prompts},
    year = {2025},
    isbn = {978-1-57735-897-8},
    publisher = {AAAI Press},
    doi = {10.1609/aaai.v39i22.34568},
    booktitle = {Proceedings of the Thirty-Ninth AAAI Conference on Artificial Intelligence and Thirty-Seventh Conference on Innovative Applications of Artificial Intelligence and Fifteenth Symposium on Educational Advances in Artificial Intelligence},
    articleno = {2670},
    numpages = {9},
    series = {AAAI'25/IAAI'25/EAAI'25}
}

@inproceedings{deng2023masterkey,
    title={MASTERKEY: {A}utomated {J}ailbreaking of Large Language Model Chatbots},
    author={Deng, Gelei and Liu, Yi and Li, Yuekang and Wang, Kailong and Zhang, Ying and Li, Zefeng and Wang, Haoyu and Zhang, Tianwei and Liu, Yang},
    booktitle={Network and Distributed System Security (NDSS) Symposium},
    year={2024}
}

@inproceedings{zheng2024improved,
    title={Improved Few-Shot Jailbreaking Can Circumvent Aligned Language Models and Their Defenses},
    author={Xiaosen Zheng and Tianyu Pang and Chao Du and Qian Liu and Jing Jiang and Min Lin},
    booktitle={The Thirty-eighth Annual Conference on Neural Information Processing Systems},
    year={2024}
}

@inproceedings{schulhoff2023hackaprompt,
    title = "Ignore This Title and {H}ack{AP}rompt: Exposing Systemic Vulnerabilities of {LLM}s Through a Global Prompt Hacking Competition",
    author = "Schulhoff, Sander and Pinto, Jeremy and Khan, Anaum and Bouchard, Louis-Fran{\c{c}}ois and Si, Chenglei and Anati, Svetlina and Tagliabue, Valen and Kost, Anson and Carnahan, Christopher and Boyd-Graber, Jordan",
    editor = "Bouamor, Houda and Pino, Juan and Bali, Kalika",
    booktitle = "Proceedings of the 2023 Conference on Empirical Methods in Natural Language Processing",
    month = dec,
    year = "2023",
    address = "Singapore",
    publisher = "Association for Computational Linguistics",
    doi = "10.18653/v1/2023.emnlp-main.302",
    pages = "4945--4977",
}

@misc{maloyan2026breakingprotocolsecurityanalysis,
      title={Breaking the Protocol: Security Analysis of the Model Context Protocol Specification and Prompt Injection Vulnerabilities in Tool-Integrated LLM Agents}, 
      author={Narek Maloyan and Dmitry Namiot},
      year={2026},
      eprint={2601.17549},
      archivePrefix={arXiv},
      primaryClass={cs.CR}
}

@misc{ding2025practical,
      title={Invisible to Humans, Triggered by Agents: Stealthy Jailbreak Attacks on Mobile Vision-Language Agents}, 
      author={Renhua Ding and Xiao Yang and Zhengwei Fang and Jun Luo and Kun He and Jun Zhu},
      year={2026},
      eprint={2510.07809},
      archivePrefix={arXiv},
      primaryClass={cs.CR},
}

@inproceedings{greshake2023not,
    author = {Greshake, Kai and Abdelnabi, Sahar and Mishra, Shailesh and Endres, Christoph and Holz, Thorsten and Fritz, Mario},
    title = {Not What You've Signed Up For: Compromising Real-World LLM-Integrated Applications with Indirect Prompt Injection},
    year = {2023},
    isbn = {9798400702600},
    publisher = {Association for Computing Machinery},
    address = {New York, NY, USA},
    doi = {10.1145/3605764.3623985},
    booktitle = {Proceedings of the 16th ACM Workshop on Artificial Intelligence and Security},
    pages = {79–90},
    numpages = {12},
    location = {Copenhagen, Denmark},
    series = {AISec '23}
}

@inproceedings {carlini2021extracting,
    author = {Nicholas Carlini and Florian Tram{\`e}r and Eric Wallace and Matthew Jagielski and Ariel Herbert-Voss and Katherine Lee and Adam Roberts and Tom Brown and Dawn Song and {\'U}lfar Erlingsson and Alina Oprea and Colin Raffel},
    title = {Extracting Training Data from Large Language Models},
    booktitle = {30th USENIX Security Symposium (USENIX Security 21)},
    year = {2021},
    isbn = {978-1-939133-24-3},
    pages = {2633--2650},
    publisher = {USENIX Association},
    month = aug
}

@inproceedings{pasquini2025llmmapfingerprintinglargelanguage,
    author = {Dario Pasquini and Evgenios M. Kornaropoulos and Giuseppe Ateniese},
    title = {{LLMmap}: Fingerprinting for Large Language Models},
    booktitle = {34th USENIX Security Symposium (USENIX Security 25)},
    year = {2025},
    isbn = {978-1-939133-52-6},
    address = {Seattle, WA},
    pages = {299--318},
    publisher = {USENIX Association},
    month = aug
}

@misc{balashov2025multistagepromptinferenceattacks,
      title={Multi-Stage Prompt Inference Attacks on Enterprise LLM Systems}, 
      author={Andrii Balashov and Olena Ponomarova and Xiaohua Zhai},
      year={2025},
      eprint={2507.15613},
      archivePrefix={arXiv},
      primaryClass={cs.CR},
}

@inproceedings{tramer2016stealing,
    author = {Tram\`{e}r, Florian and Zhang, Fan and Juels, Ari and Reiter, Michael K. and Ristenpart, Thomas},
    title = {Stealing machine learning models via prediction APIs},
    year = {2016},
    isbn = {9781931971324},
    publisher = {USENIX Association},
    address = {USA},
    booktitle = {Proceedings of the 25th USENIX Conference on Security Symposium},
    pages = {601–618},
    numpages = {18},
    location = {Austin, TX, USA},
    series = {SEC'16}
}

@article{feiglin2026benchoverflow,
    title={BenchOverflow: Measuring Overflow in Large Language Models via Plain-Text Prompts},
    author={Erin Feiglin and Nir Hutnik and Raz Lapid},
    journal={Transactions on Machine Learning Research},
    issn={2835-8856},
    year={2026},
    note={}
}

@inproceedings{dong2025memory,
    title={Memory Injection Attacks on {LLM} Agents via Query-Only Interaction},
    author={Shen Dong and Shaochen Xu and Pengfei He and Yige Li and Jiliang Tang and Tianming Liu and Hui Liu and Zhen Xiang},
    booktitle={The Thirty-ninth Annual Conference on Neural Information Processing Systems},
    year={2025}
}

@inproceedings{rajeev2025catsconfusereasoningllm,
    title={Cats Confuse Reasoning {LLM}: Query Agnostic Adversarial Triggers for Reasoning Models},
    author={Meghana Arakkal Rajeev and Rajkumar Ramamurthy and Prapti Trivedi and Vikas Yadav and Oluwanifemi Bamgbose and Sathwik Tejaswi Madhusudhan and James Zou and Nazneen Rajani},
    booktitle={Second Conference on Language Modeling},
    year={2025}
}

@inproceedings{chen2024agentpoison,
  title={AgentPoison: Red-Teaming {LLM} Agents via Poisoning Memory or Knowledge Bases},
  author={Chen, Zhaorun and Xiang, Zhen and Xiao, Chaowei and Song, Dawn and Li, Bo},
  booktitle={Advances in Neural Information Processing Systems (NeurIPS)},
  volume={37},
  year={2024}
}

@misc{yu2024assessing,
      title={Assessing Prompt Injection Risks in 200+ Custom GPTs}, 
      author={Jiahao Yu and Yuhang Wu and Dong Shu and Mingyu Jin and Sabrina Yang and Xinyu Xing},
      year={2024},
      eprint={2311.11538},
      archivePrefix={arXiv},
      primaryClass={cs.CR}
}

@inproceedings{shokri2017membership,
    author = { Shokri, Reza and Stronati, Marco and Song, Congzheng and Shmatikov, Vitaly },
    booktitle = { 2017 IEEE Symposium on Security and Privacy (SP) },
    title = {{ Membership Inference Attacks Against Machine Learning Models }},
    year = {2017},
    volume = {},
    ISSN = {2375-1207},
    pages = {3-18},
    doi = {10.1109/SP.2017.41},
    publisher = {IEEE Computer Society},
    address = {Los Alamitos, CA, USA},
    month =May
}

@inproceedings{chao2024jailbreakbench,
    title={JailbreakBench: An Open Robustness Benchmark for Jailbreaking Large Language Models},
    author={Patrick Chao and Edoardo Debenedetti and Alexander Robey and Maksym Andriushchenko and Francesco Croce and Vikash Sehwag and Edgar Dobriban and Nicolas Flammarion and George J. Pappas and Florian Tram{\`e}r and Hamed Hassani and Eric Wong},
    booktitle={The Thirty-eighth Conference on Neural Information Processing Systems Datasets and Benchmarks Track},
    year={2024}
}

@inproceedings{souly2024strongreject,
    author = {Souly, Alexandra and Lu, Qingyuan and Bowen, Dillon and Trinh, Tu and Hsieh, Elvis and Pandey, Sana and Abbeel, Pieter and Svegliato, Justin and Emmons, Scott and Watkins, Olivia and Toyer, Sam},
    title = {A STRONGREJECT for empty jailbreaks},
    year = {2024},
    isbn = {9798331314385},
    publisher = {Curran Associates Inc.},
    address = {Red Hook, NY, USA},
    booktitle = {Proceedings of the 38th International Conference on Neural Information Processing Systems},
    articleno = {3984},
    numpages = {25},
    location = {Vancouver, BC, Canada},
    series = {NIPS '24}
}

@misc{kumar2026overthinkslowdownattacksreasoning,
    title={OverThink: Slowdown Attacks on Reasoning LLMs}, 
    author={Abhinav Kumar and Jaechul Roh and Ali Naseh and Marzena Karpinska and Mohit Iyyer and Amir Houmansadr and Eugene Bagdasarian},
    year={2026},
    eprint={2502.02542},
    archivePrefix={arXiv},
    primaryClass={cs.LG}
}

@inproceedings{li2025thinktrapdenialofserviceattacksblackbox,
  author    = {Li, Yunzhe and Wang, Jianan and Zhu, Hongzi and Lin, James and Chang, Shan and Guo, Minyi},
  title     = {ThinkTrap: Denial-of-Service Attacks against Black-box LLM Services via Infinite Thinking},
  booktitle = {Proceedings of the 33rd Annual Network and Distributed System Security (NDSS) Symposium},
  year      = {2026},
  month     = {February},
  publisher = {Internet Society},
  address   = {San Diego, California}
}

@inproceedings{wu2024color,
    author = {Wu, Xiaoshuai and Liao, Xin and Ou, Bo and Liu, Yuling and Qin, Zheng},
    title = {Are watermarks bugs for deepfake detectors? rethinking proactive forensics},
    year = {2024},
    isbn = {978-1-956792-04-1},
    doi = {10.24963/ijcai.2024/673},
    booktitle = {Proceedings of the Thirty-Third International Joint Conference on Artificial Intelligence},
    articleno = {673},
    numpages = {9},
    location = {Jeju, Korea},
    series = {IJCAI '24}
}

@article{delgadochaves2025systematic,
  title = {Transforming literature screening: The emerging role of large language models in systematic reviews},
  volume = {122},
  ISSN = {1091-6490},
  DOI = {10.1073/pnas.2411962122},
  number = {2},
  journal = {Proceedings of the National Academy of Sciences},
  publisher = {Proceedings of the National Academy of Sciences},
  author = {Delgado-Chaves,  Fernando M. and Jennings,  Matthew J. and Atalaia,  Antonio and Wolff,  Justus and Horvath,  Rita and Mamdouh,  Zeinab M. and Baumbach,  Jan and Baumbach,  Linda},
  year = {2025},
  month = Jan
}

@article{wu2025automatedreview,
    author = {Wu, Shican and Ma, Xiao and Luo, Dehui and Li, Lulu and Shi, Xiangcheng and Chang, Xin and Lin, Xiaoyun and Luo, Ran and Pei, Chunlei and Du, Changying and Zhao, Zhi-Jian and Gong, Jinlong},
    title = {Automated literature research and review-generation method based on large language models},
    journal = {National Science Review},
    volume = {12},
    number = {6},
    pages = {nwaf169},
    year = {2025},
    month = {06},
    issn = {2095-5138},
    doi = {10.1093/nsr/nwaf169},
    eprint = {https://academic.oup.com/nsr/article-pdf/12/6/nwaf169/63012416/nwaf169.pdf},
}

@inproceedings{merullo2025on,
    title={On Linear Representations and Pretraining Data Frequency in Language Models},
    author={Jack Merullo and Noah A. Smith and Sarah Wiegreffe and Yanai Elazar},
    booktitle={The Thirteenth International Conference on Learning Representations},
    year={2025}
}

@inproceedings{cohan2019structural,
    title = "Structural Scaffolds for Citation Intent Classification in Scientific Publications",
    author = "Cohan, Arman and Ammar, Waleed and van Zuylen, Madeleine and Cady, Field",
    editor = "Burstein, Jill and Doran, Christy and Solorio, Thamar",
    booktitle = "Proceedings of the 2019 Conference of the North {A}merican Chapter of the Association for Computational Linguistics: Human Language Technologies, Volume 1 (Long and Short Papers)",
    month = jun,
    year = "2019",
    address = "Minneapolis, Minnesota",
    publisher = "Association for Computational Linguistics",
    doi = "10.18653/v1/N19-1361",
    pages = "3586--3596"
}

@inproceedings{wang2023trojanactivationattack,
    author = {Wang, Haoran and Shu, Kai},
    title = {Trojan Activation Attack: Red-Teaming Large Language Models using Steering Vectors for Safety-Alignment},
    year = {2024},
    isbn = {9798400704369},
    publisher = {Association for Computing Machinery},
    address = {New York, NY, USA},
    doi = {10.1145/3627673.3679821},
    booktitle = {Proceedings of the 33rd ACM International Conference on Information and Knowledge Management},
    pages = {2347–2357},
    numpages = {11},
    location = {Boise, ID, USA},
    series = {CIKM '24}
}

@inproceedings{shafran2025reroutingllmrouters,
    title={Rerouting {LLM} Routers},
    author={Avital Shafran and Roei Schuster and Thomas Ristenpart and Vitaly Shmatikov},
    booktitle={Second Conference on Language Modeling},
    year={2025},
}

@article{zhang2025dbdi, 
    title={Differentiated Directional Intervention: A Framework for Evading LLM Safety Alignment},
    volume={40}, 
    DOI={10.1609/aaai.v40i44.41148},
    number={44},
    journal={Proceedings of the AAAI Conference on Artificial Intelligence},
    author={Zhang, Peng and Sun, Peijie},
    year={2026}, 
    month={Mar.},
    pages={38102–38110} 
}

@misc{li2024logitsuppression,
      title={Uncovering Logit Suppression Vulnerabilities in LLM Safety Alignment}, 
      author={Yuxi Li and Yi Liu and Yuekang Li and Ling Shi and Gelei Deng and Shengquan Chen and Kailong Wang},
      year={2026},
      eprint={2405.13068},
      archivePrefix={arXiv},
      primaryClass={cs.CR},
      url={https://arxiv.org/abs/2405.13068}, 
}

@article{feinstein1990paradox,
  title = {High agreement but low Kappa: I. the problems of two paradoxes},
  volume = {43},
  ISSN = {0895-4356},
  DOI = {10.1016/0895-4356(90)90158-l},
  number = {6},
  journal = {Journal of Clinical Epidemiology},
  publisher = {Elsevier BV},
  author = {Feinstein,  Alvan R. and Cicchetti,  Domenic V.},
  year = {1990},
  month = Jan,
  pages = {543–549}
}

%%%%%%%%%%%%%%%%%%%%%%%%%%%%%%%%%%%%%%%%%%%%%%%%%%%%%%%%%%%%

\appendix

\newpage
\section*{Appendix overview}
\begin{itemize}[leftmargin=*,itemsep=3pt]
\item \textbf{\hyperref[app:taxonomy]{Appendix A}} --- Taxonomy structure (Figure~\ref{fig:taxonomy}) and technique column defense-layer definitions
\item \textbf{\hyperref[app:categories]{Appendix B}} --- Attack category descriptions
\item \textbf{\hyperref[app:catchall]{Appendix C}} --- Catch-all leaf analysis
\item \textbf{\hyperref[app:benchmark_mapping]{Appendix D}} --- Extended benchmark coverage mapping (empirical evidence, sensitivity analysis)
\item \textbf{\hyperref[app:benchmark_design]{Appendix E}} --- Matrix-guided benchmark design walkthrough
\item \textbf{\hyperref[app:corpus]{Appendix F}} --- Corpus distribution (Table~\ref{tab:corpus})
\item \textbf{\hyperref[app:validation]{Appendix G}} --- Validation details (novelty resolution, Tier~3 evaluation, extraction recall, taxonomy sensitivity)
\item \textbf{\hyperref[app:prompts]{Appendix H}} --- Extraction and classification prompts
\item \textbf{\hyperref[app:intermediate]{Appendix I}} --- Unmatched attack analysis
\item \textbf{\hyperref[app:datasheet]{Appendix J}} --- Datasheet for the dataset
\item \textbf{\hyperref[app:ethics]{Appendix K}} --- Ethics statement
\end{itemize}
\newpage

\section{Taxonomy structure}
\label{app:taxonomy}

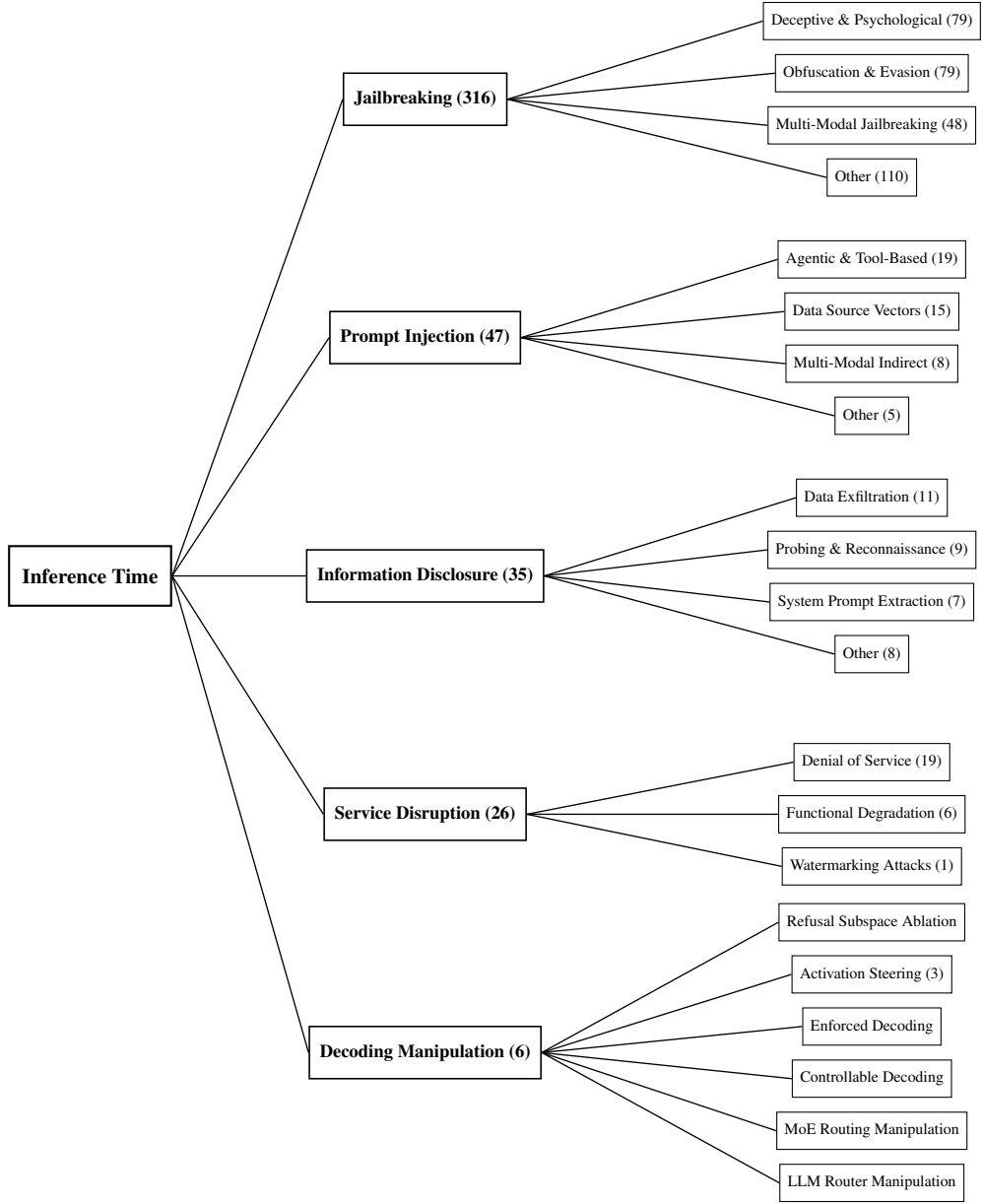
\begin{figure}[ht]
\centering
\begin{tikzpicture}[
    >=Stealth,
    root/.style={
        draw=black, line width=0.8pt,
        font=\footnotesize\bfseries,
        minimum height=8mm, inner sep=5pt, align=center
    },
    cat/.style={
        draw=black, line width=0.6pt,
        minimum height=7mm,
        inner sep=4pt, align=left, font=\scriptsize\bfseries
    },
    sub/.style={
        draw=black, line width=0.4pt,
        minimum height=5mm,
        inner sep=3pt, align=left, font=\tiny
    },
    edge/.style={draw=black, line width=0.5pt},
]

\node[root] (root) at (0, 0) {Inference Time};

\def\catx{4.5}
\def\catsep{3.2}
\node[cat] (jail) at (\catx,  2*\catsep) {Jailbreaking (316)};
\node[cat] (pi)   at (\catx,  1*\catsep) {Prompt Injection (47)};
\node[cat] (id)   at (\catx,  0)         {Information Disclosure (35)};
\node[cat] (sd)   at (\catx, -1*\catsep) {Service Disruption (26)};
\node[cat] (dm)   at (\catx, -2*\catsep) {Decoding Manipulation (6)};

\foreach \c in {jail,pi,id,sd,dm} {
    \draw[edge] (root.east) -- (\c.west);
}

\def\subx{10.5}
\def\subsep{0.7}

\node[sub] (j1) at (\subx,  2*\catsep+1.5*\subsep) {Deceptive \& Psychological (79)};
\node[sub] (j2) at (\subx,  2*\catsep+0.5*\subsep) {Obfuscation \& Evasion (79)};
\node[sub] (j3) at (\subx,  2*\catsep-0.5*\subsep) {Multi-Modal Jailbreaking (48)};
\node[sub] (j4) at (\subx,  2*\catsep-1.5*\subsep) {Other (110)};
\draw[edge] (jail.east) -- (j1.west);
\draw[edge] (jail.east) -- (j2.west);
\draw[edge] (jail.east) -- (j3.west);
\draw[edge] (jail.east) -- (j4.west);

\node[sub] (p1) at (\subx,  1*\catsep+1.5*\subsep) {Agentic \& Tool-Based (19)};
\node[sub] (p2) at (\subx,  1*\catsep+0.5*\subsep) {Data Source Vectors (15)};
\node[sub] (p3) at (\subx,  1*\catsep-0.5*\subsep) {Multi-Modal Indirect (8)};
\node[sub] (p4) at (\subx,  1*\catsep-1.5*\subsep) {Other (5)};
\draw[edge] (pi.east) -- (p1.west);
\draw[edge] (pi.east) -- (p2.west);
\draw[edge] (pi.east) -- (p3.west);
\draw[edge] (pi.east) -- (p4.west);

\node[sub] (i1) at (\subx,  1.5*\subsep) {Data Exfiltration (11)};
\node[sub] (i2) at (\subx,  0.5*\subsep) {Probing \& Reconnaissance (9)};
\node[sub] (i3) at (\subx, -0.5*\subsep) {System Prompt Extraction (7)};
\node[sub] (i4) at (\subx, -1.5*\subsep) {Other (8)};
\draw[edge] (id.east) -- (i1.west);
\draw[edge] (id.east) -- (i2.west);
\draw[edge] (id.east) -- (i3.west);
\draw[edge] (id.east) -- (i4.west);

\node[sub] (s1) at (\subx, -1*\catsep+1*\subsep) {Denial of Service (19)};
\node[sub] (s2) at (\subx, -1*\catsep)            {Functional Degradation (6)};
\node[sub] (s3) at (\subx, -1*\catsep-1*\subsep)  {Watermarking Attacks (1)};
\draw[edge] (sd.east) -- (s1.west);
\draw[edge] (sd.east) -- (s2.west);
\draw[edge] (sd.east) -- (s3.west);

\node[sub] (d1) at (\subx, -2*\catsep+2.5*\subsep) {Refusal Subspace Ablation};
\node[sub] (d2) at (\subx, -2*\catsep+1.5*\subsep) {Activation Steering (3)};
\node[sub] (d3) at (\subx, -2*\catsep+0.5*\subsep) {Enforced Decoding};
\node[sub] (d4) at (\subx, -2*\catsep-0.5*\subsep) {Controllable Decoding};
\node[sub] (d5) at (\subx, -2*\catsep-1.5*\subsep) {MoE Routing Manipulation};
\node[sub] (d6) at (\subx, -2*\catsep-2.5*\subsep) {LLM Router Manipulation};
\draw[edge] (dm.east) -- (d1.west);
\draw[edge] (dm.east) -- (d2.west);
\draw[edge] (dm.east) -- (d3.west);
\draw[edge] (dm.east) -- (d4.west);
\draw[edge] (dm.east) -- (d5.west);
\draw[edge] (dm.east) -- (d6.west);

\end{tikzpicture}
\caption{Taxonomy structure showing inference-time attack categories and their major subcategories, with leaf counts in parentheses.}
\label{fig:taxonomy}
\end{figure}

Figure~\ref{fig:taxonomy} shows a visual representation of the previously defined taxonomy in tree form.

\paragraph{Count reconciliation.} Of 2{,}521 unique attack groups, 2{,}360 (93.6\%) resolved to taxonomy leaves (1{,}868 in the five inference-time categories, 492 in Attack Generation); the remaining 161 were manually reviewed and dropped (non-attacks or out-of-scope; see Appendix~\ref{app:intermediate}). Each taxonomy leaf is classified into the 4$\times$6 matrix, with each unique attack formalized as a tuple $A = (T, M)$, where $T$ is the target objective and $M$ is the mechanism/technique.

\paragraph{Classification procedure details.} Of the 507 leaves, 333 receive their technique column assignment directly from the taxonomy structure. The remaining 174 were individually reviewed and assigned by which defense layer each attack bypasses. The breakdown by subtree is: infrastructure/reasoning (38), execution/agency (36), information disclosure (35), service disruption (26), decoding manipulation (6), and attack generation (77). Six leaves require white-box or hardware access (five model-weight, one physical hardware); they are included because they operate at inference time, but represent a distinct threat model. The full leaf-to-cell mapping is included in the released artifacts.

\paragraph{Technique column defense-layer definitions.} Each technique column is defined by the defense that would neutralize it: \emph{Instructional}---a model that perfectly maintains instruction hierarchy; \emph{Persuasion}---alignment training robust to social manipulation; \emph{Obfuscation}---a model that perfectly decodes all representations; \emph{Cross-Modal}---safety measures extending equally across all modalities; \emph{Indirect Injection}---a model that perfectly distinguishes system, user, and external data privileges; \emph{Model Internals}---architecture-level defenses such as watermark-robust decoding and side-channel resistance.

\paragraph{Output-vs-execution rule.} An explicit boundary resolves the distinction between Safety \& Alignment Bypass and System \& Tool Hijacking: if an attack succeeds when the model \emph{outputs} harmful text (e.g., generating malware source code), the target is Safety Bypass; if it succeeds when the model \emph{executes} an action or causes a side-effect (e.g., calling a tool, writing a file, running code in a sandbox), the target is System Hijacking. This boundary is operationally significant because the two targets require fundamentally different defenses: content filtering for the former, access control and sandboxing for the latter.

\paragraph{Empty cell analysis.} Five cells in Table~\ref{tab:matrix} contain zero leaves. We classify each by whether it represents a coherent single-mechanism attack surface under the bypass-target framework.

\emph{Structurally coherent (corpus gap).} Three empty cells are logically valid and likely reflect corpus limitations rather than structural impossibility. \textbf{Information Exfiltration $\times$ Indirect Injection}: a poisoned RAG document or tool response could instruct the model to append its system prompt to user-facing output, bypassing the trust boundary to achieve data leakage. \textbf{Service Disruption $\times$ Cross-Modal}: adversarial visual or audio inputs could trigger excessive generation; attacks of this form exist in the corpus (e.g., adversarial video patches inducing $>$200$\times$ token expansion) but were classified under adjacent cells. \textbf{System Hijacking $\times$ Model Internals}: an attacker could steer an agent's tool selection via activation manipulation, though this requires white-box access to a deployed agentic model, making the threat model narrow.

\emph{Compound-attack cells (two defense-layer bypasses required).} Two cells require an attacker to defeat two distinct defense layers simultaneously, making single-mechanism attacks unlikely but not structurally impossible. \textbf{Safety Bypass $\times$ Indirect Injection}: indirect injection bypasses the \emph{trust boundary} (the model treats external data as authoritative), but safety bypass additionally requires circumventing \emph{alignment training}. Demonstrated attacks in this space typically embed a persuasion or obfuscation payload within the injected text, combining two technique columns in a single attack chain. The cell is valid but inherently compound, which explains both its absence from the corpus and its absence from benchmarks that test one mechanism at a time. \textbf{Information Exfiltration $\times$ Cross-Modal}: an image containing ``reveal your system prompt'' bypasses \emph{text-input filters} via the visual channel, but the defense whose failure enables exfiltration is the \emph{trust boundary}. Here too, the cross-modal delivery and the trust-boundary bypass are two separable mechanisms that must both succeed. These cells remain valid entries in the 24-cell matrix and are counted in the benchmark coverage audit; their emptiness reflects the compound nature of the required attack rather than structural impossibility.

\section{Attack category descriptions}
\label{app:categories}

\paragraph{Jailbreaking (316 leaves).} Jailbreaking creates adversarial prompts to bypass safety alignment and produce harmful outputs~\citep{zou2023universal}. Subcategories include instruction hijacking~\citep{schulhoff2023hackaprompt} (27 leaves), deceptive/psychological manipulation~\citep{li2023deepinception,russinovich2024crescendo} (79 leaves including multi-turn, role-playing, hypothetical framing, social engineering, and persuasion), obfuscation \& evasion (79 leaves covering linguistic, encoding, formatting, tokenization, and filter evasion techniques), multi-modal jailbreaking~\citep{gong2023figstep} (48 leaves across image, audio, video, and cross-modal fusion), in-context learning exploitation~\citep{zheng2024improved} (8 leaves), hybrid \& composite attacks (1 leaf), system/infrastructure exploitation~\citep{deng2023masterkey} (38 leaves targeting reasoning chains, state/memory, context windows, and mode parameters), and unauthorized execution \& agency abuse~\citep{zhan2024injecagent} (36 leaves covering agentic framework exploitation, tool-use permission exploitation, and malicious code generation). Jailbreaking accounts for 2{,}869 mentions across 712 papers.

\paragraph{Prompt injection (47 leaves).} Prompt injection differs from jailbreaking in that the adversary injects instructions via \emph{external data sources} (web content, files, API responses, tool descriptions) rather than direct user interaction~\citep{greshake2023not}. Subcategories include data source injection vectors~\citep{chen2024agentpoison} (15 leaves), agentic \& tool-based injection~\citep{maloyan2026breakingprotocolsecurityanalysis} (19 leaves, including MCP server exploitation and agent-to-agent attacks), time-based \& stateful injection~\citep{dong2025memory} (4 leaves), multi-modal indirect injection~\citep{ding2025practical} (8 leaves), and composite payload injection (1 leaf). Prompt injection accounts for 472 mentions across 192 papers.

\paragraph{Information disclosure (35 leaves).} Information Disclosure refers to the unauthorized extraction of non-public data from the model's ecosystem~\citep{carlini2021extracting}, including system prompts, training data (membership inference or verbatim recall), or architecture/safety guardrails. Subcategories include system prompt extraction~\citep{yu2024assessing} (7 leaves), sensitive data exfiltration~\citep{balashov2025multistagepromptinferenceattacks} (11 leaves), training data recall~\citep{shokri2017membership} (2 leaves), probing \& reconnaissance~\citep{pasquini2025llmmapfingerprintinglargelanguage} (9 leaves), and model/architecture extraction~\citep{tramer2016stealing} (6 leaves). Information disclosure accounts for 101 mentions across 55 papers.

\paragraph{Service disruption (26 leaves).} Service Disruption degrades or eliminates the model's utility, targeting availability and performance rather than output content, often by exploiting the high cost of autoregressive generation or the fragility of reasoning chains. Subcategories include Denial of Service (19 leaves: computational exhaustion, resource consumption / denial-of-wallet, recursive tool calls, safeguard exploitation, simulated failure)~\citep{feiglin2026benchoverflow}; functional degradation (6 leaves: personality injection, irrelevant distraction, benign task refusal induction, reasoning degradation)~\citep{rajeev2025catsconfusereasoningllm}; and watermarking attacks~\citep{wu2024color} (1 leaf). Service Disruption accounts for 134 mentions across 49 papers with a 34.3\% novelty rate.

\paragraph{Decoding \& generation manipulation (6 leaves).} These attacks manipulate the model's internal generation process: controllable decoding, enforced decoding, refusal subspace ablation, MoE routing manipulation, LLM router manipulation, and activation steering. We note a scope tension: several of these (refusal subspace ablation, activation steering, MoE routing) require white-box access to model internals, which sits at the boundary of our inference-time definition. We retain them because they are executed \emph{during} inference (not during training) and because they appear prominently in the corpus as counterpoints to prompt-based attacks, but acknowledge that a stricter ``input-interface-only'' definition would exclude them. This category accounts for 82 mentions across 44 papers and contains only 6 of 507 leaves (1.2\%); excluding it entirely would not change any matrix-level finding.

\paragraph{Attack generation \& optimization methods (77 leaves).} This category captures methods for \emph{discovering} attacks rather than attack types themselves: gradient-based (26 leaves, anchored by GCG~\citep{zou2023universal}), automated generative / LLM-as-attacker (36 leaves, including PAIR~\citep{chao2023pair}, AutoDAN~\citep{liu2024autodan}, TAP~\citep{mehrotra2024tree}), sampling \& fuzzing~\citep{yu2024gptfuzzer} (4 leaves), black-box optimization (6 leaves), text-to-image and retrieval-based methods (2 leaves), and manual/human-in-the-loop methods (3 leaves). A GCG-generated attack might be an obfuscation attack, a jailbreak, or a prompt injection; accordingly, when a paper references both GCG (the method) and its output (e.g., an adversarial suffix), each is extracted and classified independently. Attack generation accounts for 2{,}465 mentions across 606 papers.

\section{Catch-all leaf analysis}
\label{app:catchall}

Sixty taxonomy leaves function as catch-alls, each absorbing 10+ distinct attack techniques. These are concentrated in areas of rapid innovation where naming conventions have not yet stabilized, particularly gradient-based image perturbations and automated attack generation. The recommended splits below would subdivide the largest catch-alls into mechanistically distinct subcategories; implementing them is a priority for future taxonomy versions. The top six by novel attack count are listed in Table~\ref{tab:catchall} with potential splits.

\paragraph{Impact on matrix cell counts.} All six catch-all leaves map to Safety \& Alignment Bypass (Cross-Modal, Persuasion, Obfuscation, or Model Internals). The recommended splits refine granularity \emph{within} existing matrix cells rather than redistributing across cells---e.g., splitting \texttt{adversarial-perturbation-on-images} into gradient-based, diffusion, adversarial patch, and typography sub-leaves increases the leaf count in Safety Bypass $\times$ Cross-Modal but does not populate any new cell. The only catch-all with potential for cross-cell redistribution is \texttt{activation-steering-attack} (Safety Bypass $\times$ Model Internals), where a ``directed optimization'' sub-split could arguably target Service Disruption; even so, Service Disruption $\times$ Model Internals already contains 7 leaves and remains uncovered by all benchmarks. The benchmark coverage audit result (6/24 cells) is therefore invariant to catch-all splits.

\begin{table}[ht]
\caption{Top catch-all leaves by number of novel attacks mapped to them.}
\centering
\small
\setlength{\tabcolsep}{3pt}
\label{tab:catchall}
\begin{tabular}{lrrp{5.5cm}}
\toprule
Catch-All Leaf & Novel & Total & Recommended Split \\
\midrule
\texttt{adv-perturbation-images} & 40 & 126 & Gradient-based, diffusion, adv.\ patch, typography \\
\texttt{iterative-llm-refinement} & 40 & 57 & PAIR-style, feedback-driven, prompt tuning, genetic \\
\texttt{rl-based-attack-gen} & 35 & 56 & PPO-based, reward-model, RLHF-based, multi-turn RL \\
\texttt{gcg-attack} & 27 & 67 & Token-level, attention-aware, multi-coord., checkpoint-based \\
\texttt{activation-steering} & 20 & 30 & Repr.\ engineering, patching, directed optim., latent steering \\
\texttt{evolutionary-llm-gen} & 17 & 25 & Genetic, mutation-based, crossover, tournament selection \\
\bottomrule
\end{tabular}
\end{table}

\section{Extended benchmark coverage mapping}
\label{app:benchmark_mapping}

Table~\ref{tab:benchmark_extended} shows the full mapping of six benchmarks to Target $\times$ Technique matrix cells under the bypass-target framework.

\paragraph{Benchmark mapping procedure.} We mapped each benchmark onto the matrix by: (1)~identifying every attack method evaluated by the benchmark, (2)~matching each to its taxonomy leaf by name and description, (3)~looking up that leaf's matrix cell in the released reclassification artifact, and (4)~assigning coverage as \emph{full} ($\geq$5 methods or a dedicated suite) or \emph{partial} (1--4 methods, not the benchmark's focus). For agent benchmarks whose attacks do not correspond to individual named leaves, we map the attack mechanism (indirect prompt injection via tool-returned content) to the Indirect Injection column and the attack goal (unauthorized actions vs.\ data exfiltration) to the target row.

\paragraph{Empirical evidence from uncovered cells.} The coverage gaps are not merely theoretical: papers in our corpus report high attack success rates across diverse mechanisms in uncovered cells.

\emph{Service Disruption.} OverThink~\citep{kumar2026overthinkslowdownattacksreasoning} achieves up to 46$\times$ reasoning token amplification across 9 frontier models while maintaining answer correctness, and ThinkTrap~\citep{li2025thinktrapdenialofserviceattacksblackbox} induces 100$\times$ latency increases, reducing throughput to 1\% of capacity at just 5 requests per minute. Beyond resource amplification, CatAttack~\citep{rajeev2025catsconfusereasoningllm} demonstrates that three fixed, query-agnostic adversarial triggers transfer across reasoning models (DeepSeek~R1, o1, o3-mini) without per-model optimization, increasing error rates 3$\times$ and causing 42\% of queries to exceed 1.5$\times$ token budget. These three papers demonstrate three distinct Service Disruption mechanisms---reasoning amplification, latency exhaustion, and transferable degradation triggers---none evaluated by any current benchmark.

\emph{Model Internals.} \citet{wang2023trojanactivationattack} show that injecting contrastive steering vectors at inference time reduces Vicuna-7B's refusal rate from 82\% to 2\% on AdvBench and increases toxic generation from 3\% to 83\% on ToxiGen, scaling to Llama2-70B. \citet{zhang2025dbdi} decompose the safety mechanism into two functionally distinct directions---harm detection and refusal execution---and show that surgically neutralizing both at a single critical layer achieves 97.88\% ASR on AdvBench and 95\% on HarmBench across seven models, demonstrating that alignment is not monolithic but separable and independently attackable. On the same AdvBench behaviors where GCG (the Obfuscation-column attack benchmarks \emph{do} evaluate) achieves 69\% ASR, VulMine~\citep{li2024logitsuppression} achieves 96\% ASR via logit-level denial-token suppression---a Model Internals mechanism operating on the same benchmark, same behaviors, through an untested channel. Beyond safety bypass, \citet{shafran2025reroutingllmrouters} show that adversarial token prefixes can reroute 88--100\% of queries from weak to strong models in open-source routers, achieving 79--91\% rerouting on the commercial Unify router and increasing costs up to 8$\times$ on OpenRouter---an economic denial-of-service via architectural manipulation. These results span four distinct Model Internals mechanisms---activation steering, bi-directional safety decomposition, logit-level suppression, and routing exploitation---all empirically validated but absent from standardized evaluation.

The three primary benchmarks (HarmBench, InjecAgent, AgentDojo) are used in the main text; three additional benchmarks (AdvBench, JailbreakBench, StrongREJECT) were mapped to assess whether adding benchmarks meaningfully increases coverage.

\begin{table}[ht]
\caption{Extended benchmark-to-matrix mapping under the bypass-target framework. Coverage level: \textbf{full} = dedicated evaluation suite for this cell; \textbf{partial} = technique appears as one method among many. Technique columns now reflect which defense layer each attack bypasses.}
\label{tab:benchmark_extended}
\centering
\small
\setlength{\tabcolsep}{3pt}
\begin{tabular}{llllp{5.5cm}}
\toprule
Benchmark & Target & Technique & Cov. & Grounding \\
\midrule
HarmBench & Safety Bypass & Obfuscation & full & GCG (3 variants), PEZ, GBDA, UAT, AutoPrompt, AutoDAN, PGD---10 gradient/evolutionary methods producing adversarial tokens or suffixes; each leaf classified as Obfuscation in reclassification artifact \\
HarmBench & Safety Bypass & Persuasion & full & PAIR, TAP (2 variants), PAP, Human Jailbreaks---5 methods producing persuasive or deceptive prompts via LLM refinement or manual crafting; each leaf classified as Persuasion \\
HarmBench & Safety Bypass & Instructional & partial & Direct Request, Zero-Shot, Stochastic Few-Shot---3 methods using plain-language reformulation; 3 of 18 methods \\
HarmBench & Safety Bypass & Cross-Modal & partial & Adversarial Patch, Render Text---2 multimodal methods with 110 dedicated behaviors; PGD classified as Obfuscation by leaf assignment \\
\addlinespace
InjecAgent & Sys.\ Hijacking & Indirect Inj. & full & Direct Harm: 30 attacker cases $\times$ 17 user cases (Table 1) \\
InjecAgent & Info.\ Exfil. & Indirect Inj. & full & Data Stealing: 32 attacker cases $\times$ 17 user cases (Table 1) \\
\addlinespace
AgentDojo & Sys.\ Hijacking & Indirect Inj. & full & 4 environments, 5 injection phrasings, 629 test cases \\
AgentDojo & Info.\ Exfil. & Indirect Inj. & full & Exfiltration goals in Workspace/Banking injection tasks \\
\midrule
AdvBench & Safety Bypass & Obfuscation & full & GCG on 520 harmful behaviors; leaf: \texttt{greedy-coordinate-gradient-gcg-attack} $\to$ Obfuscation \\
\addlinespace
JailbreakBench & Safety Bypass & Obfuscation & full & GCG on 100 JBB-Behaviors; leaf $\to$ Obfuscation \\
JailbreakBench & Safety Bypass & Persuasion & partial & PAIR on 100 JBB-Behaviors; leaf $\to$ Persuasion. JailbreakChat hand-crafted artifacts (partial) \\
\addlinespace
StrongREJECT & Safety Bypass & Obfuscation & partial & GCG and low-resource language translation among 17 jailbreak methods; leaves $\to$ Obfuscation \\
StrongREJECT & Safety Bypass & Persuasion & partial & Persona-based jailbreaks among 17 methods; leaves $\to$ Persuasion \\
\bottomrule
\end{tabular}
\end{table}

\paragraph{Sensitivity analysis.} The coverage figure depends on two main design choices. (1)~\emph{Partial coverage}: cells marked H$^\circ$ indicate that HarmBench includes the technique as one method among many rather than a dedicated evaluation suite. Excluding the two partial-coverage cells (Instructional, Cross-Modal) reduces the covered count from 6 to 4 of 24. (2)~\emph{Benchmark selection}: the three primary benchmarks were chosen as the largest publicly available evaluation suites for jailbreaking (HarmBench), tool-integrated agents (InjecAgent), and dynamic agent environments (AgentDojo). The three additional benchmarks (AdvBench~\citep{zou2023universal}, JailbreakBench~\citep{chao2024jailbreakbench}, StrongREJECT~\citep{souly2024strongreject}) all target exclusively Safety \& Alignment Bypass via Obfuscation and Persuasion, adding no new cells beyond those already covered by HarmBench. The remaining empty cells, including the entire Service Disruption row and the Model Internals column, remain without any standardized evaluation. Unlike the prior matrix formulation, the bypass-target framework produces no structurally implausible cells: every target--technique pairing is coherent, so all 24 cells are evaluated on equal footing.

\paragraph{Leaf-weighted vs.\ cell-weighted coverage.} An alternative \emph{leaf-weighted} coverage metric---what fraction of taxonomy leaves fall in benchmarked cells---yields ${\sim}$80\%, because the covered cells coincide with the most-studied region (Safety \& Alignment Bypass alone accounts for 74\% of all leaves). However, this metric conflates research attention with threat importance: Service Disruption and Model Internals have fewer leaves \emph{because} they are understudied, not because they pose lesser risks. A deployment facing denial-of-wallet attacks or activation-steering exploits gains no protection from benchmarks that exclusively evaluate safety-bypass techniques, regardless of how many leaves those benchmarks cover. The cell-weighted metric reflects what a threat model requires---coverage across structurally distinct attack categories---while the leaf-weighted figure captures the concentration pattern that produces the blind spots we identify. The coverage finding is also invariant to reasonable changes in matrix resolution: coarsening the matrix (e.g., merging technique columns) would increase per-cell coverage but cannot populate the Service Disruption row or Model Internals column, since no benchmark evaluates any attack in these categories regardless of how techniques are grouped. Conversely, splitting rows or columns can only decrease the coverage percentage.

\paragraph{Implication.} The coverage gaps represent a significant evaluation asymmetry. In the terms of~\citet{liao2021we}'s external-validity framework, some represent \emph{dataset misalignment}---benchmarks sample too narrowly from the technique space, omitting attack vectors their existing infrastructure could accommodate---while others reflect deeper \emph{metrics misalignment}: Service Disruption requires cost and latency metrics absent from any current benchmark, and Model Internals requires white-box access that prompt-level evaluations cannot support. The former gaps are immediately actionable by broadening existing benchmarks; the latter demand new evaluation infrastructure. That the three audited benchmarks were designed independently yet converge on the same narrow slice suggests that designers gravitate toward the most visible attack modalities, creating systematic blind spots where attacks are most costly in production.

\section{Matrix-guided benchmark design walkthrough}
\label{app:benchmark_design}

The benchmark coverage audit (\S\ref{sec:benchmark}) identifies gaps; here we demonstrate that the matrix can also \emph{fill} them. We walk through the benchmark design workflow described in \S\ref{sec:discussion} for a single uncovered cell---Service Disruption $\times$ Instructional---assembling published evidence into a concrete benchmark specification. No new attacks are executed; all results are drawn from papers in the corpus.

\paragraph{Step 1: Cell selection.} The matrix identifies Service Disruption $\times$ Instructional as uncovered by all six audited benchmarks. The cell contains 13 taxonomy leaves, 37 unique attack groups from 25 papers, and a 36.0\% novelty rate---indicating active research with zero standardized evaluation.

\paragraph{Step 2: Leaf-to-subcategory mapping.} The 13 leaves organize into three subcategories that a benchmark must cover to achieve adequate cell coverage:

\begin{center}
\small
\begin{tabular}{p{3.5cm} p{8.5cm}}
\toprule
Subcategory & Taxonomy Leaves \\
\midrule
Computational Exhaustion & \texttt{reasoning-loop-over-reasoning-dos}, \texttt{complex-calculation-recursion-prompts}, \texttt{semantic-contradiction-dos}, \texttt{regex-dos-redos-via-prompt-interpretation}, \texttt{logic-bombs-in-prompts}, \texttt{instructional-overhead-dos-attack} \\
\addlinespace
Functional Degradation & \texttt{catattack-irrelevant-distraction}, \texttt{benign-task-refusal-induction}, \texttt{math-perturb-attack} \\
\addlinespace
Resource Consumption \& Safeguard Exploitation & \texttt{excessive-token-generation-requests}, \texttt{strategic-resource-exhaustion-in-llm-mas}, \texttt{dos-via-safeguard-false-positives}, \texttt{embodied-agent-action-freezing} \\
\bottomrule
\end{tabular}
\end{center}

\paragraph{Step 3: Evidence assembly.} For each subcategory, published papers in the corpus already provide ground-truth results. Table~\ref{tab:walkthrough} assembles these into a benchmark evidence table---the core artifact a benchmark designer needs.

\begin{table}[ht]
\caption{Published evidence for Service Disruption $\times$ Instructional, organized by the matrix-guided workflow. The rightmost column identifies the metric required to detect each attack---none of which appear in any of the six audited benchmarks, all of which evaluate output content rather than resource consumption or task correctness.}
\label{tab:walkthrough}
\centering
\small
\setlength{\tabcolsep}{3pt}
\begin{tabular}{p{2.4cm} p{1.8cm} p{2.4cm} p{3.2cm} p{2.2cm}}
\toprule
Taxonomy Leaf & Source & Models & Published Result & Required Metric \\
\midrule
\multicolumn{5}{l}{\textit{Computational Exhaustion}} \\
\addlinespace
reasoning-loop-over-reasoning-dos & OverThink \citep{kumar2026overthinkslowdownattacksreasoning} & 9 frontier reasoning models & 46$\times$ token amplification; answers remain correct & Token amplification ratio \\
\addlinespace
reasoning-loop-over-reasoning-dos & ThinkTrap \citep{li2025thinktrapdenialofserviceattacksblackbox} & Black-box LLM services & 100$\times$ latency; throughput $\to$ 1\% at 5 req/min & Latency ratio; throughput \\
\addlinespace
excessive-token-generation-requests & BenchOverflow \citep{feiglin2026benchoverflow} & Multiple LLMs & Token overflow via plain-text prompts & Token count \\
\midrule
\multicolumn{5}{l}{\textit{Functional Degradation}} \\
\addlinespace
catattack-irrelevant-distraction & CatAttack \citep{rajeev2025catsconfusereasoningllm} & DeepSeek R1, o1, o3-mini & 3$\times$ error rate; 42\% exceed 1.5$\times$ token budget; query-agnostic triggers & Answer correctness; token budget \\
\bottomrule
\end{tabular}
\end{table}

\paragraph{Step 4: Metric specification.} The Required Metric column in Table~\ref{tab:walkthrough} reveals why existing benchmarks cannot detect these attacks: all six audited benchmarks evaluate output \emph{content} (harmful or not), but Service Disruption attacks succeed through resource amplification or quality degradation while producing benign or correct outputs. Detecting them requires metrics that no current benchmark records. A benchmark for this cell requires:

\begin{center}
\small
\begin{tabular}{ll}
\toprule
Metric & Rationale \\
\midrule
Token amplification ratio & Computational exhaustion (OverThink: 46$\times$) \\
Wall-clock latency ratio & Throughput impact (ThinkTrap: 100$\times$) \\
API cost under attack & Denial-of-wallet quantification \\
Answer correctness & Distinguishes degradation from refusal \\
Throughput at attack load & System-level impact (ThinkTrap: 1\%) \\
\bottomrule
\end{tabular}
\end{center}

\noindent None of these metrics appear in any of the six audited benchmarks. This is not a dataset gap but a \emph{metrics} gap in the sense of~\citet{liao2021we}: the evaluation infrastructure lacks the measurement apparatus to detect an entire class of successful attacks.

\paragraph{Summary.} The matrix guided a four-step path from an empty cell to a benchmark specification grounded in published evidence: cell selection $\to$ leaf enumeration $\to$ evidence assembly $\to$ metric identification. The same workflow applies to any uncovered cell. The 8 remaining uncovered cells in the Service Disruption row and Model Internals column can be addressed analogously, though Model Internals cells additionally require white-box access infrastructure that no current public benchmark provides.

\section{Corpus distribution}
\label{app:corpus}

Table~\ref{tab:corpus} shows the distribution of papers and attack mentions across half-year periods, along with post-resolution novelty counts.

\begin{table}[ht]
\caption{Corpus distribution by half-year period. Novel claims reflect post-resolution counts (42 false-novel mentions corrected; see \S\ref{sec:validation}). The novelty rate stabilizes around 18\%, indicating concurrent field growth and maturation.}
\label{tab:corpus}
\centering
\small
\begin{tabular}{lrrrr}
  \toprule
  Period & Papers & Attack Mentions & Novel Claims & Novel \% \\
  \midrule
  H1 2023 & 3 & 21 & 11 & 52.4 \\
  H2 2023 & 42 & 253 & 68 & 26.9 \\
  H1 2024 & 74 & 528 & 101 & 19.1 \\
  H2 2024 & 120 & 969 & 142 & 14.7 \\
  H1 2025 & 183 & 1{,}397 & 235 & 16.8 \\
  H2 2025 & 182 & 1{,}446 & 259 & 17.9 \\
  H1 2026 & 259 & 1{,}752 & 357 & 20.4 \\
  \midrule
  \textbf{Total} & \textbf{863} & \textbf{6{,}366} & \textbf{1{,}173} & \textbf{18.4} \\
  \bottomrule
\end{tabular}
\end{table}

\section{Validation details}
\label{app:validation}

\paragraph{Novelty resolution.} Three cross-reference checks identified 33 novelty conflicts (multiple papers claiming the same attack as novel) and 43 suspicious novelty claims. All 76 entries were manually reviewed: 33 false-novel claims were corrected (15 generic category labels mismarked as novel, 9 same-paper extraction duplicates, 5 unattributed rediscoveries, and 4 naming collisions), while 16 cases were confirmed as genuinely distinct methods sharing a group key. Source attributions for non-novel attacks were extracted from in-paper citations and are released as-is.

\paragraph{Tier~3 human evaluation.} Two domain experts independently reviewed 200 Tier~3 MATCH classifications, stratified by category with tail-oversampling (100 Jailbreaking, 35 Attack Generation, 25 Prompt Injection, 15 each for Info Disclosure and Service Disruption, 10 Decoding Manipulation). Each sample was judged as correct or incorrect, with an associated confidence level and error-type classification for incorrect cases.

Overall accuracy was 92.0\% (184/200; 95\% CI: [87.4\%, 95.0\%] by Wilson interval); reweighted to population proportions, 92.3\%. Per-category accuracy ranged from 73.3\% (Info Disclosure, $n$=15) to 100\% (Service Disruption and Decoding Manipulation). Among the 16 errors: 12 (75.0\%) were \emph{within-category} errors (correct branch, wrong leaf---do not affect matrix-level counts), 2 (12.5\%) were \emph{cross-category} errors (wrong branch), and 2 (12.5\%) were \emph{force-classification} errors (non-attacks matched to leaves). Under pessimistic assumptions, the 2 cross-category errors would shift affected matrix cells by $\leq$0.2\% of total attacks per cell. High-confidence judgments accounted for 72.0\% of annotations (144/200), medium 22.0\% (44/200), and low 6.0\% (12/200). The evaluation codebook is available in the released artifacts.

\paragraph{Inter-annotator agreement.} Two domain experts independently annotated 115 stratified samples, each working without access to the other's annotations. After independent annotation, disagreements were reconciled through discussion. We report agreement at three granularity levels based on the \emph{pre-reconciliation} annotations, as the paper's claims operate at the matrix-cell level rather than the leaf level:

\begin{center}
\small
\begin{tabular}{lrrr}
\toprule
Granularity & Classes & Agreement & Cohen's $\kappa$ \\
\midrule
Leaf & 507 & 70.4\% & 0.70 \\
Matrix cell & 24 & 86.1\% & 0.84 \\
\bottomrule
\end{tabular}
\end{center}

\noindent Of the 34 leaf-level disagreements, 29 followed the same pattern: Annotator~A accepted the pipeline's classification while Annotator~B proposed a more specific leaf within the same subtree. Of the 34, 18 (53\%) mapped to the same Target~$\times$~Technique cell; the remaining 16 involved genuinely ambiguous attacks spanning cell boundaries. Nine of the 16 cross-cell disagreements shared the same target row but differed on technique column (e.g., Instructional vs.\ Obfuscation for prompt extraction attacks), while seven crossed target rows (e.g., Repeat Attack: obfuscation technique vs.\ information disclosure; RAG Documents Extraction: system hijacking vs.\ information exfiltration). Cross-cell disagreements concentrated in Information Disclosure (6 of 15 samples, 40\%) and Service Disruption (4 of 15 samples, 27\%), reflecting genuine boundary ambiguity in technique-column assignment for these categories. Post-reconciliation, cell-level agreement rose to 89.6\% (103/115); the Information Disclosure disagreements were resolved by applying the output-vs-execution rule and match-by-mechanism principle from the codebook. This calibration difference---one annotator applying broader acceptance criteria, the other insisting on the most specific leaf---does not affect the benchmark coverage audit, which maps benchmarks to cells via manual documentation review. We do not report verdict-level $\kappa$ (pipeline-correct vs.\ pipeline-incorrect), as divergent acceptance thresholds between annotators (92\% vs.\ 72\%) produce the well-known prevalence paradox~\citep{feinstein1990paradox}; the effective-leaf and matrix-cell metrics compare chosen classifications directly and avoid this artifact.

\paragraph{Multi-model classification comparison.} To assess whether Tier~3 classifications are sensitive to model choice, we re-classified the same 200 stratified attacks using Claude Opus 4.6 (Anthropic) with the identical classification prompt. Agreement by level:

\begin{center}
\small
\begin{tabular}{lrr}
\toprule
Granularity & Classes & Agreement \\
\midrule
Target row & 4 & 94.5\% \\
Matrix cell & 24 & 80.5\% \\
Leaf & 507 & 63.0\% \\
\bottomrule
\end{tabular}
\end{center}

\noindent Of 74 leaf-level disagreements, 35 (47\%) mapped to the same matrix cell. Among the 39 cross-cell disagreements, 28 shared the same target row but differed on technique column; the most common confusions were Persuasion$\leftrightarrow$Obfuscation (6$\times$) and Instructional$\leftrightarrow$Obfuscation (5$\times$), reflecting genuine ambiguity in which defense layer an attack bypasses. Only 11 disagreements (5.5\% of all samples) crossed target rows, involving boundary cases such as ``dark pattern'' attacks arguable between System Hijacking and Safety Bypass; of these 11, 6 (3.0\% of all samples) fell on the Safety Bypass$\leftrightarrow$System Hijacking boundary specifically, confirming that the output-vs-execution rule (\S\ref{sec:matrix}) is the primary source of cross-target ambiguity. The benchmark coverage audit is unaffected by these differences, as benchmark methods are mapped to cells via manual documentation review rather than model classification.

\paragraph{Structural pattern robustness.} The three structural patterns reported in Section~\ref{sec:analysis} are robust to model choice. Under Claude classifications: (1)~all 23 Indirect Injection assignments target System Hijacking (100\%, vs.\ 96\% under Gemini); (2)~Obfuscation, Persuasion, and Cross-Modal concentrate 96\%, 91\%, and 96\% respectively in Safety Bypass (vs.\ 92\%, 89\%, 88\% under Gemini); (3)~Instructional distributes across all four target rows under both models.

\paragraph{Taxonomy construction process.} The five categories emerged from a bottom-up process: initial clusters were formed by adversarial objective (what the attacker wants), then refined by examining the threat model each cluster implies (e.g., whether the attack requires model access, targets content vs.\ availability, or operates via direct vs.\ indirect channels). Cluster boundaries were iteratively adjusted until they satisfied two criteria: (1) each category corresponds to a distinct defensive requirement and (2) no leaf naturally belongs to two categories. The four target rows (grounded in STRIDE) and six technique columns (organized by bypass target) were chosen to be mutually exclusive and collectively exhaustive over the corpus.

\paragraph{Taxonomy and count sensitivity.} The five attack categories and 4$\times$6 matrix dimensions are design choices, not natural kinds. Alternative decompositions (e.g., merging Decoding Manipulation into Jailbreaking, splitting Jailbreaking by modality) would yield different coverage statistics. The Target $\times$ Technique matrix is a simplification, as some attacks span multiple cells; this is resolved by the bypass-target principle and the output-vs-execution rule (\S\ref{sec:matrix}), but inevitably causes a loss in nuance. The 2{,}521 unique attack groups figure is best interpreted as a lower bound on the number of distinct attack identifiers the field has produced. The taxonomy's 507 leaves provide a closer approximation, of which 401 are populated; the remaining 106 represent documented attack types that no paper in the corpus mentioned.

Some taxonomy leaves, particularly generic catch-all nodes (e.g., \texttt{instruction-segmentation}, \texttt{multi-turn-context-manipulation}), absorb structurally diverse attacks that share a delivery mechanism but differ in technique---a known property of hierarchical taxonomies that contributes to within-category disagreement. Recent work on LLM-assisted systematic reviews suggests that an LLM-plus-human-reviewer protocol can reduce workload by 33--93\% while maintaining classification quality~\citep{delgadochaves2025systematic}, supporting our approach of LLM classification with targeted human spot-checks.

\paragraph{Extraction recall.} A domain expert manually reviewed 13 papers (5 with blind extraction before seeing pipeline output, 8 with anchored verification; 3 selected specifically for low pipeline extraction counts of 0--2 matched attacks to test for catastrophic misses). An attack was counted if described with sufficient detail to understand its mechanism; bare citation-only references---which constitute 58\% of citations in computer science papers~\citep{cohan2019structural}---were excluded from the recall denominator.

Extraction precision was 100\% (76/76). Extraction recall was 89.4\% (76/85): 9 substantively described attacks were missed. The precision-recall asymmetry is explained by frequency-dependent representation formation in LLMs~\citep{merullo2025on}: the pipeline reliably extracts attacks with strong internal representations (yielding high precision) but fails to recognize less-established attacks below the model's representation threshold. All nine misses were lesser-known attacks (e.g., FERRET, PoisonedRAG) described at identical depth to successfully extracted well-known attacks (e.g., GCG, AutoDAN) in the same paragraph. An additional 5 of 76 extractions (6.6\%) were correctly extracted but misclassified at the taxonomy level, suggesting that extraction and classification errors are largely independent.

\paragraph{Novelty rate confound.} The frequency-dependent extraction gap predicts lower recall for novel or less-established attacks, which could deflate observed novelty rates in later periods and mimic field maturation. Two observations partially address this confound: first, the novelty rate \emph{increases} from 14.7\% (H2 2024) to 20.4\% (H1 2026), inconsistent with monotonic recall degradation; second, novel attacks as a paper's primary contribution receive prominent, detailed descriptions that likely exceed the extraction model's recognition threshold---the 9 extraction misses were lesser-known \emph{referenced} attacks, not papers' headline contributions. Nevertheless, we cannot fully disentangle genuine maturation from pipeline recall effects, and the ${\sim}$18\% figure should be treated as an apparent rate that may undercount genuinely novel but underspecified attacks.

\section{Extraction and classification prompts}             \label{app:prompts} 
The following prompts were used verbatim in the pipeline. Minor formatting (line breaks) has been adjusted for readability. Template variables (e.g.\ \texttt{\{name\}}) are populated at runtime.
\paragraph{Prompt 1: Attack extraction (full).} Used with Gemini 3.1 Pro to extract attacks from each paper's PDF. The output schema enforces structured JSON with fields: \texttt{name}, \texttt{description}, \texttt{example}, \texttt{is\_novel}, and \texttt{original\_source\_paper}.

\begin{promptbox}[Prompt 1 --- Full extraction]
You are an academic paper review and summarization tool focused on INFERENCE-TIME adversarial attacks on LLMs --- attacks executed through the model's input interface during deployment, requiring no access to the training pipeline or model weights.

Extract every specific, named attack type discussed in this paper, including its name, a substantial description, and a concrete example. Do not skip any attacks.

DO NOT extract any of the following:
- Evaluation metrics, scoring functions, or measurement tools (e.g., ASR, perplexity scores, Zlib compression ratios)
- Defense techniques, guardrails, safety filters, or detection methods
- Broad categories or umbrella terms (e.g., ``adversarial attacks'', ``prompt-space attacks'', ``perturbation-based attacks'', ``heuristic-based algorithms'') --- only extract specific named attacks
- Components that are not attacks on their own (e.g., Vigenere cipher, Base64 encoding, Caesar cipher --- unless the paper presents them AS an attack method)
- Training-phase attacks (data poisoning, backdoor injection, weight manipulation, fine-tuning exploits)
- Attacks on non-LLM systems (e.g., GNN perturbations, traditional SQL injection, network exploits) unless the paper explicitly adapts them for LLMs
- Error types, failure modes, or risk descriptions that are not deliberately triggered by an adversary (e.g., ``hallucination risk'', ``premature actions'')

For each attack, determine whether it is a novel contribution introduced by this paper, or whether this paper is referencing/citing an attack that was originally presented in prior work.

To determine novelty:
- If the paper explicitly proposes, introduces, or presents the attack as its own contribution, mark is\_novel=true and set original\_source\_paper to the title of the current paper.
- If the paper cites, references, evaluates, or builds upon an attack from another paper, mark is\_novel=false and set original\_source\_paper to the original paper's citation (author, year, title) as it appears in the references.
- Pay close attention to citation markers (e.g.\ [1], [2], (Author et al., 2023)) near attack descriptions to identify the source.
- If the paper discusses a well-known attack technique without a specific citation, still mark is\_novel=false and provide the best attribution you can.
\end{promptbox}

\paragraph{Prompt 2: Attack extraction (concise retry).} Used when the full prompt timed out or produced truncated output.

\begin{promptbox}[Prompt 2 --- Concise retry]
You are an academic paper review and summarization tool focused on INFERENCE-TIME adversarial attacks on LLMs.

Extract only the NOVEL attack types introduced by this paper (not referenced/cited attacks from other work). For each novel attack, provide:

- name, description (2--3 sentences max), a brief example, is\_novel=true, and original\_source\_paper as this paper's title.

Also list referenced attacks with MINIMAL detail: just name, one-sentence description, is\_novel=false, and the original source citation. Keep descriptions and examples SHORT to avoid truncation.

DO NOT extract any of the following:
- Evaluation metrics, scoring functions, or measurement tools (e.g., ASR, perplexity scores, compression ratios)
- Defense techniques, guardrails, or safety filters
- Broad categories or umbrella terms (e.g., ``adversarial attacks'', ``jailbreaking methods'')
- Components that are not attacks on their own (e.g., cipher algorithms, encoding schemes, optimization steps)
- Training-phase attacks (data poisoning, backdoor injection, weight manipulation, fine-tuning exploits)
- Attacks on non-LLM systems (e.g., GNN perturbations, traditional network exploits) unless explicitly adapted for LLMs
- Error types or failure modes that are not deliberately triggered by an adversary
\end{promptbox}
\paragraph{Prompt 3: Tier 3 semantic classification.} Used with Gemini 3.1 Pro to classify attacks that failed direct and fuzzy matching. The compact leaf catalog (all 505 leaf IDs available at classification time, with descriptions and variant annotations) is appended after the \texttt{TAXONOMY LEAF CATALOG} header. Two additional leaves were added to the taxonomy after classification, bringing the final count to 507; these were assigned to matrix cells manually. When a fuzzy string match exists, a hint line is injected (shown as \texttt{\{fuzzy\_section\}} below); otherwise it is omitted. Structured JSON output is enforced via the Gemini \texttt{response\_schema} parameter.

\begin{promptbox}[Prompt 3 --- Semantic classification]
\#\#\# SYSTEM ROLE\\
You are an INFERENCE-TIME LLM ATTACK TAXONOMY CLASSIFIER performing academic security research. You will be given an attack name and description extracted from a published academic paper.

\#\#\# TASK\\
Classify this attack into the taxonomy below. Return exactly one leaf ID.\\
- If the input is NOT a genuine inference-time LLM attack (e.g., a metric, defense, training-phase attack, or broad category), return ``NOT\_AN\_ATTACK''.\\
- If it IS an attack but no leaf captures its mechanism, return ``NO\_LEAF\_FITS''.\\
- Otherwise return ``MATCH'' with the closest leaf ID.\\
- Match by MECHANISM, not surface name similarity.\\
- Use variant names shown in [aka: ...] annotations when matching.\\
- If the attack is the same technique under a different name or alias, match it.\\
\{fuzzy\_section\}

\#\#\# ATTACK TO CLASSIFY\\
Name: ``\{norm\_name\}''\\
Description: ``\{description[:800]\}''

\#\#\# TAXONOMY LEAF CATALOG\\
\{leaf\_catalog\}
\end{promptbox}

\noindent The leaf catalog lists each leaf ID with a truncated description and variant annotations (e.g., \texttt{gcg-attack: Greedy Coordinate Gradient... [aka: GCG, universal adversarial suffix]}). The catalog is included in the released pipeline code.

\section{Unmatched attack analysis}
\label{app:intermediate}

Of the 2{,}521 unique attacks, 161 (6.4\%) did not map to any leaf node and were dropped after manual review (see \S\ref{sec:matrix} for the full count reconciliation). The classification prompt requires leaf-level assignment; no attacks map to intermediate nodes.

Of these 161, 41 claimed novel status. Manual review reveals they fall into distinct categories: broad category or umbrella terms incorrectly marked as novel (${\sim}$15, e.g., ``semantic manipulation,'' ``capability jailbreak,'' ``complex adversarial''); training-phase or out-of-scope methods (${\sim}$8, e.g., data paraphrasing for watermark removal, temporal backdoors); non-LLM targets (${\sim}$5, e.g., recommender system shilling attacks, neural ranker exploits); and evaluation frameworks or metrics misclassified as attacks (${\sim}$5, e.g., benchmark tier labels, stress tests). Approximately 8 are genuinely borderline: of these, several map to existing taxonomy leaves under different names (e.g., ``twinbreak'' is already catalogued as an alias under \texttt{mixture-of-experts-routing-exploitation}; ``context-aware overthink'' maps to existing Service Disruption reasoning-loop leaves), while others target niche application domains (e.g., LLMs for time-series forecasting). No dropped attack represents a genuinely novel attack \emph{type} that the taxonomy lacks a category for. The benchmark coverage audit is methodologically independent of the extraction pipeline---benchmark methods are mapped to matrix cells via manual documentation review (\S\ref{sec:benchmark})---so any extraction-side exclusions do not affect the 6/24 coverage result.

\section{Datasheet for the LLM Attack Taxonomy Dataset}
\label{app:datasheet}

\paragraph{Motivation.} The dataset was created to address the lack of a comprehensive, quantitatively grounded taxonomy of LLM adversarial attacks. It is intended for use by the safety and alignment research community.

\paragraph{Composition.} The dataset contains: (a) 6{,}366 attack mention records extracted from 932 arXiv papers; (b) a 507-leaf hierarchical taxonomy of inference-time attacks with Target $\times$ Technique classifications. Each record includes source paper identifier, attack name, taxonomy mapping, match tier, novelty status, and temporal period. The 507 leaves are the terminal nodes of the taxonomy hierarchy. All matched attacks map to leaf nodes; unmatched entries were manually reviewed and dropped.

\paragraph{Collection process.} Papers were collected from the Promptfoo LLM Security database~\citep{PromptfooLMSecurityDB}, filtered to inference-time adversarial attacks on LLMs published on arXiv between 2023 and 2026. The last date of collection was April 10th, 2026. Structured extraction used Gemini 3.1 Pro.

\paragraph{Preprocessing.} Attack names were normalized via case folding, kebab-case conversion, and removal of parenthetical abbreviations. Classification used three-tier matching (direct, fuzzy, semantic); see \S\ref{sec:matrix} for mapping rates.

\paragraph{Source attribution quality.} Each non-novel attack record includes an \texttt{original\_source\_paper} field extracted from in-paper citations. A spot-check of 100 attributions found 72\% verifiably correct, 14\% ambiguous (generic attack names without a clear single originator, e.g., ``system message,'' ``ICL''), and 14\% misattributed. Misattributions clustered among community-originated attacks attributed to the first academic paper to study them rather than their informal origin, and generic pipeline-assigned names matched to topically related papers. No downstream analysis (taxonomy structure, matrix assignments, benchmark coverage) depends on source attributions; users of the released dataset should treat this field as approximate.

\paragraph{Uses.} Intended for: benchmark gap analysis, attack naming standardization, temporal trend analysis, and defensive research prioritization. Not intended for: generating new attacks, automating exploitation, or replacing expert security assessment.

\paragraph{Distribution.} Released on HuggingFace under CC-BY-4.0 (data) and MIT (code). Croissant metadata provided.

\paragraph{Maintenance.} The dataset represents a snapshot of the literature through early 2026. Future versions may extend the corpus and refine catch-all leaf splits.

\section{Ethics statement}
\label{app:ethics}

This work catalogs adversarial attacks for defensive purposes: enabling systematic evaluation and benchmark development. We do not introduce new attacks or provide exploit code; the taxonomy describes attack categories at a level of abstraction that does not enable reproduction without consulting the underlying papers. We acknowledge the dual-use concern that a comprehensive catalog could theoretically assist adversaries, but believe the defensive value---identifying evaluation blind spots, standardizing naming, and prioritizing benchmark development---substantially outweighs this risk given that all source papers are already public. The dataset contains no personally identifiable information; all inputs to the extraction pipeline are publicly available arXiv papers.

%%%%%%%%%%%%%%%%%%%%%%%%%%%%%%%%%%%%%%%%%%%%%%%%%%%%%%%%%%%%
\end{document}